\documentclass[sigconf]{acmart}

\usepackage{bm}
\usepackage[linesnumbered,ruled,vlined]{algorithm2e}
\usepackage{multirow}
\usepackage{stmaryrd}

\usepackage{amsmath,amssymb}
\usepackage{graphicx}
\usepackage{subfigure}

\newtheorem{theorem}{Theorem}

\newtheorem{lemma}{Lemma}

\newtheorem{definition}{Definition}
\newtheorem{problem}{Problem}
\newtheorem{prune}{Prune Rule}
\newcommand{\tabincell}[2]{\begin{tabular}{@{}#1@{}}#2\end{tabular}}

\AtBeginDocument{%
  \providecommand\BibTeX{{%
    \normalfont B\kern-0.5em{\scshape i\kern-0.25em b}\kern-0.8em\TeX}}}

\copyrightyear{2022}
\acmYear{2022}
\setcopyright{acmcopyright}\acmConference[WWW '22]{Proceedings of the ACM Web Conference 2022}{April 25--29, 2022}{Virtual Event, Lyon, France}
\acmBooktitle{Proceedings of the ACM Web Conference 2022 (WWW '22), April 25--29, 2022, Virtual Event, Lyon, France}
\acmPrice{15.00}
\acmDOI{10.1145/3485447.3512198}
\acmISBN{978-1-4503-9096-5/22/04}
\acmSubmissionID{w09fp1965}

\begin{document}

%%
%% The "title" command has an optional parameter,
%% allowing the author to define a "short title" to be used in page headers.
\title{Listing Maximal $k$-Plexes in Large Real-World Graphs}

%%
%% The "author" command and its associated commands are used to define
%% the authors and their affiliations.
%% Of note is the shared affiliation of the first two authors, and the
%% "authornote" and "authornotemark" commands
%% used to denote shared contribution to the research.

\author{Zhengren Wang}
\affiliation{%
  \institution{University of Electronic Science and Technology of China}
  \city{Chengdu}
  \country{China}}
\email{zr-wang@outlook.com}

\author{Yi Zhou}
\authornote{Corresponding author.}
\affiliation{%
  \institution{University of Electronic Science and Technology of China}
  \city{Chengdu}
  \country{China}}
\email{zhou.yi@uestc.edu.cn}

\author{Mingyu Xiao}
\affiliation{%
  \institution{University of Electronic Science and Technology of China}
  \city{Chengdu}
  \country{China}}
\email{myxiao@uestc.edu.cn}

\author{Bakhadyr Khoussainov}
\affiliation{%
  \institution{University of Electronic Science and Technology of China}
  \city{Chengdu}
  \country{China}}
\email{bmk@uestc.edu.cn}

%%
%% By default, the full list of authors will be used in the page
%% headers. Often, this list is too long, and will overlap
%% other information printed in the page headers. This command allows
%% the author to define a more concise list
%% of authors' names for this purpose.
\renewcommand{\shortauthors}{Z. Wang, Y. Zhou, M. Xiao, and B. Khoussainov.}

%%
%% The abstract is a short summary of the work to be presented in the
%% article.
\begin{abstract}
Listing dense subgraphs in large graphs plays a key task in varieties of network analysis applications like community detection.
Clique, as the densest model, has been widely investigated.
However, in practice, communities rarely form as cliques for various reasons, e.g., data noise.
Therefore, $k$-plex, -- graph with each vertex adjacent to all but at most $k$ vertices,
is introduced as a relaxed version of clique.
Often, to better simulate cohesive communities, an emphasis is placed on connected $k$-plexes with small $k$.
In this paper, we continue the research line of listing all maximal $k$-plexes and maximal $k$-plexes of prescribed size.
Our first contribution is algorithm \emph{ListPlex} that lists all maximal $k$-plexes in $O^*(\gamma^D)$ time for each constant $k$, where $\gamma$ is a value related to $k$ but strictly smaller than 2,
and $D$ is the degeneracy of the graph that is far less than the vertex number $n$ in real-word graphs.
Compared to the trivial bound of $2^n$, the improvement is significant, and our bound is better than all previously known results.
In practice, we further use several techniques to accelerate listing $k$-plexes of a given size, such as structural-based prune rules, cache-efficient data structures, and parallel techniques.
All these together result in a very practical algorithm.
Empirical results show that our approach outperforms the state-of-the-art solutions by up to orders of magnitude.
\end{abstract}

%%
%% The code below is generated by the tool at http://dl.acm.org/ccs.cfm.
%% Please copy and paste the code instead of the example below.
%%


%%
%% Keywords. The author(s) should pick words that accurately describe
%% the work being presented. Separate the keywords with commas.


\maketitle

\section{Introduction}
\label{sec_introduction}

\subsection{Motivation}
Finding \emph{cohesive groups} (or \emph{communities}) has received a lot of attention from various areas such as social network analysis and web mining, and is also a fundamental problem in graph algorithms.
The community can be modeled in many ways. For example, the notion of \emph{clique} is the strictest and arguably the most studied community model. A clique is a subgraph in which vertices are pairwise connected, i.e., a complete subgraph. A large body of literature dedicated to related problems has emerged, e.g., enumerating cliques in graphs \cite{xiao2017exact}, sparse graphs \cite{chang2013fast,eppstein2011listing}, uncertain graphs \cite{mukherjee2016enumeration}, limited main memory \cite{cheng2012fast}, and optimizing the running time as a function of the output \cite{conte2016sublinear}.
The clique model has also been applied in many domains such as data mining \cite{cheng2012fast}, bio-informatics \cite{butenko2006clique} and ad-hoc wireless network \cite{chen2004clustering}.

In real-world graphs, due to various reasons such as the existence of data noise, communities rarely appear in the form of cliques \cite{balasundaram2011clique,conte2017fast,conte2018d2k}.
Therefore, other forms of \emph{relaxed cliques} are proposed as relaxations of the notion of clique.
For example, the \emph{$k$-core} \cite{cheng2011efficient} relaxes the vertex degree, \emph{$k$-club} \cite{pajouh20162} relaxes pairwise distance of vertices and \emph{$k$-clique densest subgraph} \cite{tsourakakis2015k} relaxes density of induced subgraph.
In this paper, we continue on this line of research by focusing on $k$-\emph{plex}, -- the notion that has been receiving increasing attention and popularity in recent years \cite{zhou2020enumerating,conte2018d2k,xiao2017fast,conte2017fast}.

A $k$-plex is a relaxed clique model first proposed in \cite{seidman1978graph}.
A $k$-plex is a graph in which each vertex's degree is at least $n-k$, where $n$ is  the number of vertices in the graph.
In other words, a $k$-plex allows every vertex missing at most $k$ links to other vertices (including itself) compared to the clique.
Note that a $1$-plex is just a clique.
%\textcolor{blue}{postulate that all our $k$-plexes will be large connected graphs}.
A $k$-plex in a graph is called \emph{maximal} if and only if it is not a subgraph of any larger $k$-plex.

\paragraph{Listing maximal $k$-plexes}
In this paper, we will study the problem of listing all maximal $k$-plexes from a given graph. It would seem that the listing of maximal $k$-plexes will be also useful in applications where maximal clique listing is applied.
Additionally, the $k$-plex listing has other potential applications like link prediction.

From the theoretical point of view, listing all maximal $k$-plexes is hard.
In fact, for any given $k$, this problem is NP-hard \cite{balasundaram2011clique} and it is known that the number of maximal $k$-plexes is exponential in the worst-case \cite{moon1965cliques}.
Therefore, a large number of existing studies focus on the design of practically efficient methods.
%A \textcolor{blue}{large} number of maximal $k$-plex enumeration algorithms have been proposed.
The majority of these algorithms have been derived and motivated by the \emph{Bron-Kerbosch} algorithm \cite{Bron:1973:AFC:362342.362367}, though it was originally designed to only list maximal cliques.
\citeauthor{wu2007parallel} (\citeyear{wu2007parallel}) adapted the Bron-Kerbosch algorithm to list maximal $k$-plexes with a few new rules to prune unnecessary searches \cite{wu2007parallel}.
\citeauthor{wang2017parallelizing} (\citeyear{wang2017parallelizing}) integrated more heuristic pruning rules and applied multi-thread parallelization technique \cite{wang2017parallelizing}.
\citeauthor{zhou2020enumerating} (\citeyear{zhou2020enumerating}) devised a novel branch heuristic with a worst-case time complexity proof.
With their branch heuristic, the running time of Bron-Kerbosch algorithm is improved from $O^*(2^n)$ to $O^*(\gamma_k^n)$ where  $\gamma_k$ is related to $k$ but strictly smaller than $2$ \footnote{The notation $O^*$ omits the polynomial factors.}.
 Aside from the Bron-Kerbosch variants, there is another type of algorithms which have bounded \emph{delay} between the output of two consecutive solutions.
\citeauthor{berlowitz2015efficient} (\citeyear{zhou2020enumerating}) initialized such kind of study by providing a polynomial-time delay algorithm for the problem \cite{berlowitz2015efficient,cohen2008generating}.

\paragraph{Listing large maximal $k$-plexes}
We also study the problem of listing large maximal $k$-plexes, i.e., listing maximal $k$-plexes which have at least $l$ vertices, $l$ being a large number, say at least $2k-1$.
The problem was originally proposed to amend two issues that arise in modeling the communities by maximal $k$-plexes \cite{conte2017fast,conte2021meta,conte2018d2k,zhou2020enumerating}.
First, it is observed that there are enormous maximal $k$-plexes in real-world graphs, and empirically most of them are small or even unconnected.
However, in community detection application, communities should be large and densely connected subgraphs.
Second, existing maximal $k$-plex listing algorithms can only handle graphs with hundreds to thousands of vertices in days.
But large sparse graphs are ubiquitous these days, e.g., the \emph{webbase-2001} web-graph has more than a hundred million vertices and more than a billion edges \cite{BoVWFI}).
%Fortunately, there is a simple solution to amend such issues -- we can recast community detection problem as listing all maximal $k$-plexes which have at least $l$ vertices, $l$ being at least $2k-1$.

Fortunately,  by requiring that the output $k$-plexes must be at least larger than a threshold  $l$ ($l\ge 2k-1$), the two issues can be alleviated.
Due to the structural property of $k$-plexes (Property 3 in \cite{xiao2017fast}), a $k$-plex with at least $2k-1$ vertices is densely connected, i.e., the shortest length of paths between every two vertices is not larger than 2. Therefore, the first issue does not exist.
For the second issue, with a lower bound requirement on the size of output $k$-plexes, the  performance of listing algorithm can be also accelerated with many powerful strategies \cite{conte2017fast,conte2021meta,conte2018d2k}. For instance, in \cite{conte2017fast,conte2021meta}, \citeauthor{conte2017fast} (\citeyear{conte2017fast}) took advantage of the size constraint to remove a large portion of unfruitful vertices from the input graph, which made \citeauthor{berlowitz2015efficient}'s listing algorithm possible to run on graphs of millions of vertices.
\citeauthor{conte2018d2k} (\citeyear{conte2018d2k}) further used decomposition and parallel techniques, leading to a listing algorithm capable of running on some web-scale graphs, e.g., the \emph{it-2004} graph.
\citeauthor{zhou2020enumerating} (\citeyear{zhou2020enumerating}) used the same decomposition framework as in \cite{conte2018d2k} so that their Bron-Kerbosch pivot heuristic can accommodate large real-world graphs.

\vspace{-2mm}

\subsection{Contributions}
Motivated by the aforementioned studies, we develop the most efficient algorithm for listing both maximal $k$-plexes and large maximal $k$-plexes from sparse real-world graphs.

\smallskip

{\bf 1.}  {\em We propose ListPlex, an algorithm that lists all maximal $k$-plexes with provably worst-case running time.}
    The general idea of ListPlex is a marriage of new decomposition scheme and an efficient Bron-Kerboch search.
    Our analysis discloses that for each constant $k$, ListPlex has a worst-case time bound $O^*(\gamma_k^D)$ where $D$ is the degeneracy number of the input graph and $\gamma_k$ is related to $k$ but strictly smaller than 2.
    As far as we know, it is the first algorithm that reduces the exponent of running time from $n$ to $D$.
    Due to the power-law distribution of most real-world graphs, $D\ll n$ in most cases, e.g., the webbase-2001 has more than a thousand million vertices but its degeneracy number is only 1506. To some extent, this bound provides theoretical evidence for the good performance of our algorithm.

\smallskip

{\bf 2.} {\em We optimize the practical performance of ListPlex for listing large maximal $k$-plexes of size at least $2k-1$.}
    It is known that listing large maximal $k$-plexes is of more real-world importance than purely listing all maximal $k$-plexes. Thus, we study efficient implementation techniques from multiple perspectives.
    From algorithmic perspective, we suggest strong prune rules to reduce the search space of our algorithm.  From the computational system perspective, we propose new data structures to reduce cache misses and increase parallelism.
    All optimization techniques bring evident speedup for processing large sparse real-world graphs.

Our experiments %extensive empirical study
show that ListPlex outperforms the state-of-the-art approaches in terms of both problems.
For example, our parallel algorithm can list all large maximal $2$-plexes (with $l=800$) from the huge webbase-2001 graph with over one billion edges in 1 minutes. This is almost an order of magnitude speedup compared to the best-known parallel approach.

All codes are available at \url{https://github.com/joey001/ListPlex.git}.

%{\color{red} Existing work in finding the maximum $k$-plexes:\cite{seidman1983network}, \cite{balasundaram2011clique}[Branch and Cut] \cite{xiao2017fast},\cite{gao2018exact}[Branch and Bound] \\Enumerating maximal $k$-plexes: \cite{conte2017fast}[reduce], \cite{wang2017parallelizing}[Parallel], \cite{berlowitz2015efficient}[polynomial delay],\cite{conte2018d2k}[D2K] }

%In real-life tasks such as community detection, we are more interested in $k$-plexes of non-trivial size and bounded diameter. For example, An independent set size at most $k$ also forms a legal $k$-plex, which, however, cannot represent a community. In the latest research in \cite{conte2018d2k} who constraint the target $k$-plexes of size at least $l$ and diameter bounded by 2. Indeed, the diameter can be bounded by a more compact way. We only need to set $l>2k-2$ for a $l>0$. Due to the conclusion founded in \cite{xiao2017fast}, the $k$-plexes of size large than $l$ enjoy the properties of being connected and more importantly, has a diameter at most 2.

%The scalability must be considered for the design of practical algorithms. The real network is often massive, with up to millions or even billions of vertices and edges, which hinders the application of many existing algorithms in reality. Fortunately, these graphs often enjoy features such as sparse and pow-law degree distribution.  In this paper, we make use of the fact that large real-life graph often has low core number to design scalable algorithm.

\section{Backgrounds}
\subsection{Basic notations}
\label{sec_prelimiaries}
Let $G = (V, E)$ be a simple and undirected input graph,  where $V$ and $E$ are the sets of vertices and edges, respectively.
We will let $n=|V|$ and $m=|E|$ in this paper.
For $v\in V$ and a positive integer $k$, we use $N^k_G(v)$ to denote the set of vertices with distance exactly $k$ to $v$ in $G$. The vertices in $N^k_G(v)$ are also called \emph{$k$-hop neighbors} of $v$. The set $N^1_G(v)$ may be simply written as $N_G(v)$ and 1-hop neighbors may be simply called \emph{neighbors}. The degree of a vertex $v$ is $|N_G(v)|$. The maximum degree among all vertices in $G$ is denoted by $\Delta$.
When the underlying graph $G$ is clear from the content, we may ignore the subscript $G$ and write $N^k_G(v)$ as $N^k(v)$.
Given a vertex set $P \subseteq V$, let $G[P]$ be the \emph{subgraph induced by $P$}.
The \emph{diameter} of $G$ is the maximum distance among all pairs of vertices in $G$.

A permutation of vertices $v_1\dots v_n$ is called a \emph{degeneracy ordering} (or \emph{core ordering}) of the graph $G$ if for each $i$, vertex  $v_i$ has the minimum degree in the induced subgraph $G[\{v_i,...,v_n\}]$.
The degeneracy ordering of a graph can be computed in linear time by the algorithm that repeatedly removes a node with the minimum degree until the graph becomes empty \cite{batagelj2003m}.
For a degeneracy ordering $\eta = v_1\dots v_n$, the degree of $v_i$ in $G[\{v_i,...,v_n\}]$ is called the \emph{core number of $v_i$}.
It is known that for any degeneracy ordering of the same graph, the largest core number among all vertices is the same and is called as \emph{degeneracy} (or core number). We denote it by $D$.

%That is, the degeneracy number is independent of the ordering and we use $D$ to denote the degeneracy of the graph.

Let $\eta$ be an ordering of $V$, say $\eta = v_1$,$\ldots $,$v_n$. For any two different vertices $v_i$ and $v_j$, denote $v_i\prec_\eta v_j$  if $v_i$ precedes $v_j$ in $\eta$, otherwise $v_i\succ_\eta v_j$.
For any $v_i$, let $N_{\prec_\eta}(v_i)$ be $N(v_i)\cap \{v_1,...,v_{i-1}\}$, $N_{\succ_\eta}(v_i)$ be $N(v_i)\cap \{v_{i+1},...,v_n\}$. Similarly, let $N^2_{\prec_\eta}(v_i)$ denotes $N^2(v_i)\cap \{v_1,...,v_{i-1}\}$ and $N^2_{\succ_\eta}(v_i)$ denotes $N^2(v_i) \cap \{v_{i+1},...,v_n\}$.

As defined above, a $k$-plex is a graph such that each vertex is not adjacent to at most $k$ vertices (including itself) in the graph. Thus, a 1-plex is a clique, i.e., a complete graph.
A subgraph $G'$ of $G$ is called a \emph{maximal $k$-plex} if  $G'$ is not a subgraph of any larger $k$-plex.
So a maximal $k$-plex is always an induced subgraph.
In this paper, we are interested in listing all maximal $k$-plexes of a graph.

\begin{problem}[Listing maximal $k$-plexes]
\label{problem_diameter2}
Given a graph $G=(V,E)$, a positive integer $k$, list all maximal $k$-plexes of $G$.
\end{problem}

\subsection{Some properties}
We present basic properties of $k$-plexes. These are important for our algorithm design. Proofs of the lemmas below as well as missing proofs in the rest of the paper are left in the Appendix.

\begin{lemma}
Any induced subgraph of a $k$-plex is still a $k$-plex.
\end{lemma}
This property is known in the literature~\cite{cohen2008generating,zhou2021improving,seidman1978graph}. It will be frequently used in our algorithm.
For example, we can validate the maximality of a $k$-plex, i.e., a $k$-plex $G[P]$ is maximal if there is no vertex that can be added into $G[P]$ such that $G[P]$ is still a $k$-plex.

\begin{lemma}\label{connectedP}
Any $k$-plex with at least $2k-1$ vertices is a connected graph with the diameter at most $2$. A $k$-plex with at most $2k-2$ vertices may be disconnected.
\end{lemma}

%Xiao et al.~\cite{} also further show that a connected $k$-plex with more $2k-c$ vertices has a diameter at most $c$ for any $c\geq 2$.

Lemma~\ref{connectedP} is also known in the literature~\cite{xiao2017exact,conte2017fast}.  It shows $2k-1$ is a key  boundary between connectedness and unconnectedness.

In practice,  the $k$-plex is closely related to the community detection problem which asks for dense and large communities from a large network \cite{conte2018d2k,zhu2020community}.
Using maximal $k$-plex as a graph model of the community, we translate the community detection as listing  maximal $k$-plexes that are at least connected, and with prescribed number of vertices.
By Lemma~\ref{connectedP}, any $k$-plex of size at least $2k-1$ must be connected and even diameter-2 bounded.
Therefore, it is rational to form the practical community detection as finding all maximal $k$-plexes of size at least $l$, where $l$ is a given lower bound value and $l$ must be at least $2k-1$.

\begin{problem}[Listing large maximal $k$-plexes]
\label{problem_q}
Given a graph $G=(V,E)$, two positive integers $k$ and $l$ where $l \ge 2k-1$, list all maximal $k$-plexes with at least $l$ vertices.
\end{problem}

% It will be helpful for us to understand the difference of the new proposed algorithm.
\subsection{Existing Bron-Kerbosch based algorithms}
%\subsection{Enumerating all maximal k-plexes}
Before we present our algorithm, we introduce the Bron-Kerbosch algorithm and its variants as they are closely related to ours.
%algorithm.

\subsubsection{The fundamental Bron-Kerbosch Algorithm}
Many existing algorithms for listing maximal $k$-plexes, as in  \cite{wu2007parallel,wang2017parallelizing,bentert2018listing}, stem from the Bron-Kerbosch algorithm that was originally designed from listing maximal cliques \cite{Bron:1973:AFC:362342.362367,cheng2012fast}.
We review the main idea of the Bron-Kerbosch algorithm for listing $k$-plexes.

The algorithm is recursive. We leave the pseudo-code in Alg. \ref{alg_BK} in the Appendix. It calls a recursive procedure \emph{BKRec} with three disjoint sets as parameters, i.e., $P$, $C$ and $X$. $P$ represents the set of vertices that should be contained in the $k$-plex in the current stage. $C$ includes the remaining \emph{candidate vertices} for enumerating. $X$ contains \emph{excluded vertices}. %which have been processed.
They are excluded from the $k$-plex to avoid non-maximal solutions.

BKRec lists all maximal $k$-plexes $G[P']$ satisfying the following three properties: \
 (i) $P \subseteq P'$, \  (ii) $P' \subseteq P\cup C$, and  (iii) $\forall v\in X$, the subgraph $G[\{v\} \cup P']$ is not a $k$-plex.

Given a graph $G=(V,E)$ and an integer $k>0$, the algorithm calls BKRec initialized with $P=X=\emptyset$ and $C =V$. Then the algorithm iteratively branches on a vertex in $C$ by including it to either $P$
or $X$. We will use BKPlex$(G,k)$ to denote this algorithm.

\paragraph{Complexity} As mentioned in \cite{zhou2020enumerating}, the Bron-Kerbosch requires $O^*(2^n)$ time in the worst-case, where $n$ is the number of vertices in the input graph.
Although several pruning rules were suggested for the Bron-Kerbosch in \cite{wu2007parallel,wang2017parallelizing}, but the worst-case running time bound was not improved. 
%\textcolor{green}{Wu and Pei (\citeyear{wu2007parallel}) improved the Bron-Kerbosch algorithm with a few new rules to prune unnecessary searches.
%Wang et al. (\citeyear{wang2017parallelizing}) integrated more heuristic pruning rules and applied multi-thread parallelization technique.[may be duplicate with the introduction?]} But the worst-case running time bound was not improved.

\subsubsection{\citeauthor{zhou2020enumerating}'s Pivot Heuristic}

\citeauthor{zhou2020enumerating} (\citeyear{zhou2020enumerating}) improved the Bron-Kerbosch algorithm with a \emph{pivot} heuristic \cite{zhou2020enumerating}.
They observed that for any graph $G$, either $G$ is a $k$-plex or there is a vertex $v$ not adjacent to at least $k+1$ vertices in $G$, including itself. As such, they designed a pivot heuristic which always branches on the vertex of minimum degree in the graph. 

\paragraph{Complexity} The pivot heuristic can reduce the total number of branches and then improve the worst-case running time from $O^*(2^n)$ to  $O^*(\gamma_k^n)$, where $\gamma_k$ is a number related to $k$ but strictly smaller than 2.

%Our proposed algorithm will also embed this algorithm as a subprocess. So we leave the details of the pivot heuristic in the remaining part of the paper.

\subsubsection{\citeauthor{conte2018d2k}'s Decomposition Algorithm}
In \cite{conte2018d2k}, \citeauthor{conte2018d2k} proposed a decomposition-based algorithm, namely D2K, for listing $k$-plexes with the diameter at most 2.
%Generally, for each vertex, D2K builds a subgraph and then adopts Bron-Kerbosch algorithm to list all the maximal $k$-plexes that contains such a vertex in the subgraph. This may reduce the running time greatly in practice.
D2K first sorts the vertices of $G$ by degeneracy ordering $v_1$, $\dots$, $v_n$.
Then, for each $v_i$, D2K builds a subgraph $G_i=(V_i,E_i)$ induced by $\{v_i\}\cup N_{\succ_\eta}(v_i)\cup N^2_{\succ_\eta}(v_i)$.
The Bron-Kerbosch algorithm is then called to search all maximal $k$-plexes in $G_i$.
However, a maximal $k$-plex $G_i[P]$ of $G_i$ is not a maximal $k$-plex of the original graph if a vertex preceding $v_i$ can form a larger $k$-plex with $P$. Hence, for every maximal $k$-plex $G_i[P]$ emitted by the Bron-Kerbosch search algorithm, D2K further validates that no other vertex in ${v_1,\dots,v_{i-1}}$ can form a $k$-plex with $P$ before outputting it.

\paragraph{Complexity}
D2K restricts the search space to each subgraph $G_i=(V_i,E_i)$ and so the search size is bounded by $O^*(\sum_{i}2^{|V_i|})$.
Recall that $D$ is the degeneracy of the input graph $G$ and $\Delta$ is the maximum degree of $G$. It holds that $|V_i|\leq D\Delta$ for each $i$. Thus, $\sum_{i}2^{|V_i|}\leq n2^{D\Delta}$.
Due to the sparsity of many real-world graphs, $\Delta$ and $D$ are normally small values. The algorithm thus performs better than the Bron-Kerbosch algorithm in these large graphs.

\section{Listing All Maximal $k$-Plexes}
We present our algorithm, ListPlex, for listing all maximal $k$-plexes.

\subsection{The main Structure}

%Lemma \ref{connectedP} indicates that the size $2k-1$ is an important boundary between connectedness and unconnectedness.
%We are inspired by the this observation and propose the following idea to listing maximal $k$-plexs without repetition.
%For small maximal $k$-plexes, say with size at most $2k-2$, we can enumerate them directly by brute force.
%In fact, usually the parameter $k$ is also small and most previous algorithms only tested the cases of $k\leq 6$.
%As $k$ can be seem as a small constant value, the enumerating time is polynomially bounded.
%%For the other cases, we consider $k$-plexes as relaxations of cliques. So they should be dense and connected at least. It is reasonable to restrict the size of $k$-plexes to at least $2k-1$. Furthermore, for small $k$-plexes, say with size at most $2k-2$, usually we can enumerate them directly by brute force since $k$ is not big in real life problems.
%As for the remaining larger $k$-plexes, we know that these $k$-plexes are connected graphs with the diameter at most 2.
%Following the idea of Conte et al.'s decomposition algorithm, we can list maximal $k$-plexes containing a vertex $v$ in the subgraph with $v$ and its 2-hop neighbors.
%The framework is shown in Algorithm~\ref{}.
%As listing  maximal $k$-plexes of size at least $2k-1$ vertices, in the rest of this section, we will focus on listing  maximal $k$-plexes of size at least $2k-1$ vertices.

Our algorithm contains two parts that are to list maximal $k$-plexes of size at most $2k-2$ vertices and at least $2k-1$ vertices, respectively.
As mentioned in Lemma~\ref{connectedP}, maximal $k$-plexes of size at most $2k-2$ may not be connected and this kind of $k$-plex is not interesting in practice.
In fact, usually the parameter $k$ is also small and most previous algorithms only tested the cases of $k\leq 5$.
In our algorithm, we will modify the Bron-Kerbosch algorithm by adding the size constraint to find all maximal $k$-plexes of size at most $2k-2$.

Next, we will focus on listing  maximal $k$-plexes of size at least $2k-1$.
By Lemma~\ref{connectedP}, we know that maximal $k$-plexes of size at least $2k-1$ are connected graphs with the diameter at most 2.
So following the idea of Conte et al.'s decomposition algorithm, we list maximal $k$-plexes containing a vertex $v_i$ by only considering the local subgraph induced by $\{v_i\}\cup N_{\succ_\eta}(v_i)\cup N^2_{\succ_\eta}(v_i)$.
However, we further use some techniques to reduce the search space again and get a significantly improved running time bound.
%We still use some techniques to accelerate this part. Our algorithm uses a top-down idea and calls a procedure $Check(G')$ on some subgraphs.
%The procedure $Check(G')$ will check whether $G'$ is a maximal $k$-plex and whether it has been

\subsection{Listing maximal $k$-plexes larger than $2k-2$}
In this subsection, we focus on listing all maximal $k$-plexes of size at least $2k-1$.
The pseudo-code corresponds to the second part in Alg. \ref{alg_ListPlex}. We will explain the idea and each step of the algorithm.

%LargePlex takes a graph $G=(V,E)$ and an integer $k$ as the input.
First, ListPlex sorts the $V$ by a degeneracy ordering $\eta=v_1\dots v_n$.
From $v_1$ to $v_n$, ListPlex iteratively lists \emph{maximal $v_i$-leaded $k$-plexes with at least $2k-1$ vertices}.

\begin{definition}
Given an ordering $\eta=v_1\dots v_n$ of the vertices of $G$, a $v_i$-leaded $k$-plex is a $k$-plex $G[S]$ such that $v_i$ is in $S$ and it holds that
$v_i\prec_\eta u$ for each vertex $u\in S\setminus\{v_i\}$.
 A $v_i$-leaded $k$-plex is maximal if it is not a subgraph of any $k$-plex in $G[\{v_i,...,v_n\}]$.
\end{definition}
Note that a maximal $v_i$-leaded $k$-plex may not be maximal in the original graph $G$. So in our algorithm, when a maximal $v_i$-leaded $k$-plex is found, we also check its maximality in $G$.
%In lines 5 to 13 in Alg. \ref{alg_ListPlex}, the algorithm lists all the \emph{maximal $v_i$-leaded $k$-plex} and check its maximality in $G$.

The core part of the algorithm is to find all maximal $v_i$-leaded $k$-plexes. Instead of using a brute force method, we dramatically reduce the search space by utilizing the structural properties. %of a local subgraphs.

\begin{lemma}
\label{lemma_diam2}
Given an ordering $\eta$ of the vertices of $G$, let $G[P]$ be a $v_i$-leaded $k$-plex induced by $P$ and $|P|\ge 2k-1$.
Then $G[P]$ must be a subgraph of $G[\{v_i\}\cup N_{\succ_\eta}(v_i)\cup N^2_{\succ_\eta}(v_i)]$. Furthermore, $P$ contains at most $k-1$ vertices from $N^2_{\succ_\eta}(v_i)$.
\end{lemma}

Let us call $G_i=G[\{v_i\}\cup N_{\succ_\eta}(v_i)\cup N^2_{\succ_\eta}(v_i)]$ as the \emph{seed graph of $v_i$}.
%Inspired by Lemma~\ref{lemma_diam2}, ListPlex first enumerates all subsets $S\subseteq N^2_{\succ_\eta}(v_i)$ with size $|S|\le k-1$, and then use a procedure called BKPivot to find all maximal $k$-plexes that contain $S$.
Given a $v_i$ and a subset $S\subseteq N^2_{\succ_\eta}(v_i)$ such that $|S|\le k-1$, let us call $P_s=\{v_i\}\cup S$ as a \emph{seed set}.
%By Lemma~\ref{lemma_diam2}, for each seed set, we call BKPivot to  search $k$-plexes containting $P_s$ from $G_i$.
For set  $P_s=\{v_i\}\cup S$,  by Lemma \ref{lemma_diam2}, we call BKPivot to search maximal $v_i$-leaded $k$-plexes that must contain $P_s$.
The elaboration of BKPivot is left to the next subsection.
In the current stage, we specify that for each seed set $P_s=\{v_i\}\cup S$, BKPivot emits all maximal $k$-plexes that must include $P_s$, possibly include some vertices in $N_{\succ_\eta}(v_i)$ and  must not include vertices in $N^2_{\succ_\eta}(v_i)\setminus S$.
%This step will reduce the candidate set from vertex set of $G_i$ to only $N_{\succ_\eta}(v_i)$ and accelerate the algorithm in practice.

For each maximal $v_i$-leaded $k$-plex $G[P]$ found by BKPivot, ListPlex further tests the maximality of $G[P]$ in the input graph $G$. That is to say, if a vertex in $N_{\prec_\eta}(v_i)$ and $N^2_{\prec_\eta}(v_i)$ can form a larger $k$-plex with $P$, then $G[P]$ is not maximal in $G$. Otherwise, $G[P]$ is maximal and $P$ is emitted.

% \textcolor{red}{Recall that decomposing the input graph $G$ into $G_i$ reduces the search space from $|V|$ to $D\Delta$. The key point of ListPlex is, by utilizing the structural properties and building seed sets, ListPlex further excludes $N^2_{\succ_\eta}(v_i)$ from candidates and dramatically reduces the search space again.[I don't think this paragraph is necessary!]}
\begin{algorithm}[ht!]
    % \scriptsize
    \DontPrintSemicolon
        \caption{Our maximal $k$-plex listing algorithm}
      %  \KwIn{a graph $G=(V,E)$, an integer $k>1$}
       % \KwOut{All vertex sets that induce maximal $k$-plexes in $G$. }
        \label{alg_ListPlex}
        \emph{ListPlex}$(G,k)$\\
        \Begin{
            \textbf{Part I:}
            Use the basic Bron-Kerbosch algorithm to list all the maximal $k$-plexes of size at most $2k-2$.

            \textbf{Part II:}\\
            Sort $V$ by degeneracy ordering as $\{v_1,...,v_n\}$\\
            \For{$i\gets 1,...,n$}{
                %$ \gets  N_{>}(v_i)\cup N^2_{>}(v_i)$\\
                %$L \gets N_{<}(v_i)\cup N^2_{<}(v_i)$\\
                %$B_i\gets G[L,R]$ \\
                Build seed graph $G_i= G[\{v_i\}\cup  N_{\succ_\eta}(v_i) \cup N^2_{\succ_\eta}(v_i)]$ \\
                \For{any $S\subseteq N^2_{\succ_\eta}(v_i) $ that $|S|\le k-1$}{
                    Build seed set $P_s=\{v_i\}\cup S, C_s \gets N_{\succ_\eta}(v_i), X_s \gets N^2_{\succ_\eta}(v_i)\setminus S$\\
                    Call BKPivot($G_i, k, P_s, C_s, X_s$)\\
                    \For{ each {$P$} emitted by BKPivot }
                    {
                            %CheckMaximality($G_i, B_i, Plex$)
                            \If{$|P|> 2k-2$ and $\nexists u\in N_{\prec_\eta}(v_i)\cup N^2_{\prec_\eta}(v_i)$ that $G[\{u\}\cup P]$ is a $k$-plex in $G$}{
                                emit $P$
                            }
                    }

                }
            }

        }

\end{algorithm}

\begin{figure*}[htbp]
\centering  %å›¾ç‰‡å…¨å±€å±…ä¸­
\subfigure[Graph decomposition.]{
    \label{Fig.sub.1}
    \includegraphics[width=0.4\textwidth]{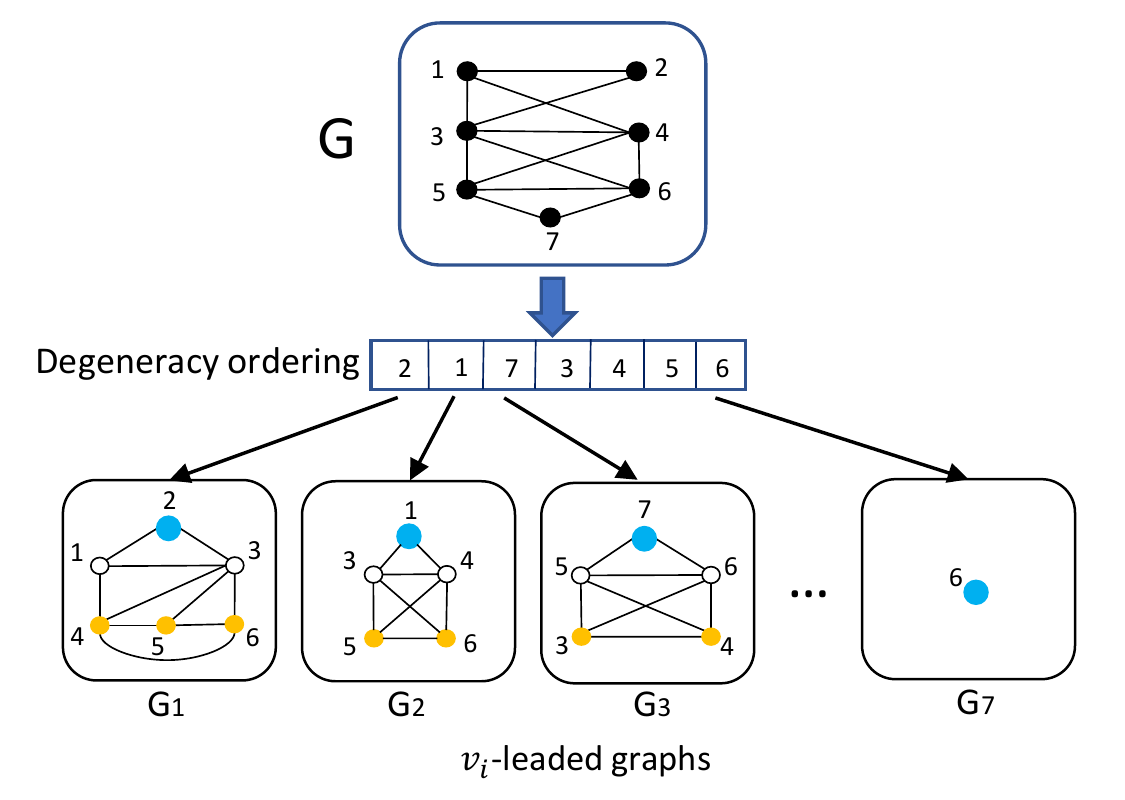}
}
\subfigure[Build seeds and call BKPivot.]{
    \label{Fig.sub.2}
    \includegraphics[width=0.4\textwidth]{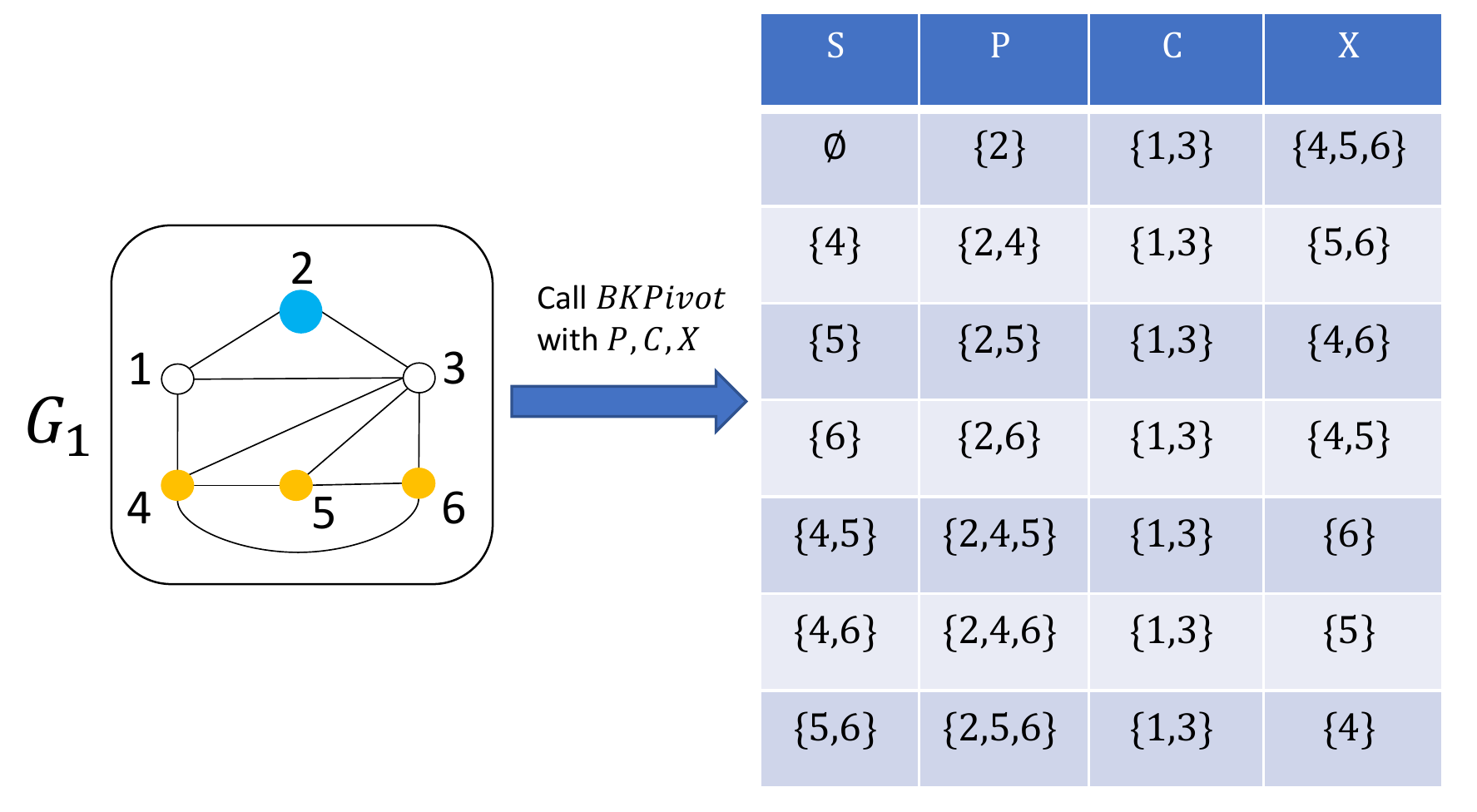}
}
\caption{An example of the ListPlex algorithm.
(a) sort $V$ in degeneracy ordering $\eta$ and induce seed graphs $G_i$ for each $v_i \in \eta$.
(b) enumerate $S\subseteq N^2_{\succ_\eta}(v_i)$ with bound $|S|\le k-1\;(k=3)$ and call BKPivot with $P_s, C_s, X_s$.
}
\label{Fig-lstplex-example}
\end{figure*}

\subsection{The BKPivot algorithm}
We introduce BKPivot. 
It is also a branching algorithm following the style of the basic Bron-Kerbosch algorithm and it
accepts three disjoint sets $P$, $C$ and $X$ playing the same roles as those in the Bron-Kerbosch algorithm.
% So BKPivot lists maximal $k$-plexes containing all vertices in $P$ and no vertex in $X$. 
However, it additionally integrates some ideas into its branch scheme to reduce more vertices. 

\begin{figure}[htb]
  \centering
  \includegraphics[width=1.0\columnwidth]{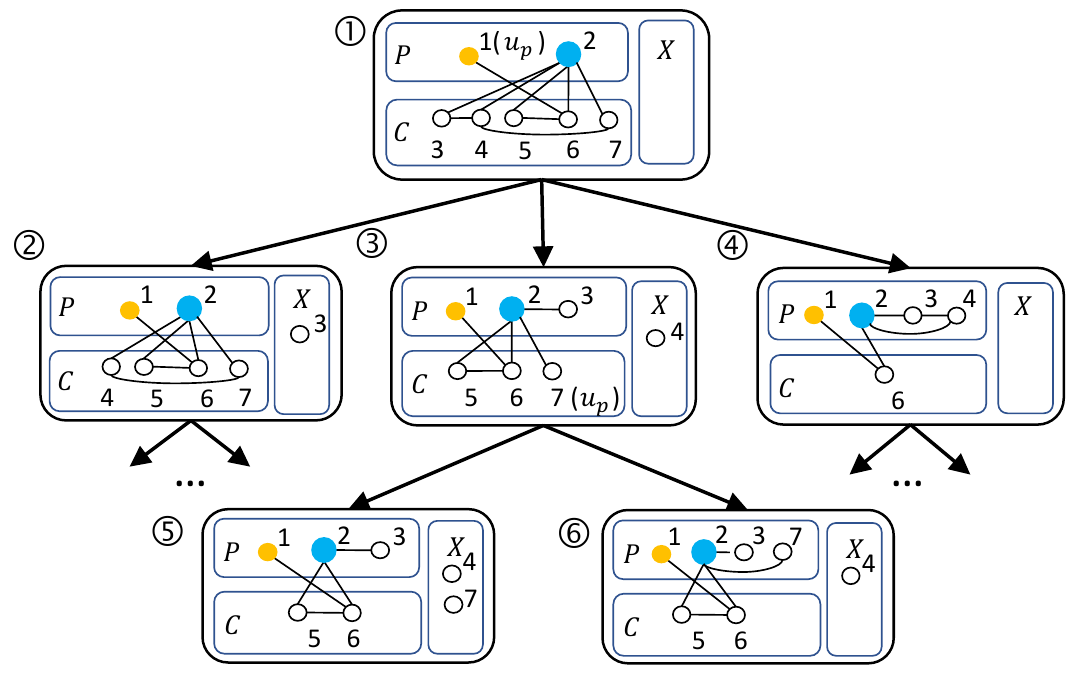}\\
  \caption{An example of BKPivot's branch scheme with $k=4$.
    In node 1, pivot $u_p=1 \in P$, $P\setminus N(u_p)=\{1,2\}$ and $C\setminus N(u_p)=\{3,4,5,7\}$. At most $k'=k-|P\setminus N(u_p)|=2$ non-neighbors of $u_p$ can be moved from $C$ to $P$. Node 1 generates three branches, i.e., node 2, node 3 and node 4.
  In node 3, there are several vertices of minimum degree. Assume pivot $u_p=7 \in C$, node 3 further generates two branches by moving $u_p$ to $X$ or $P$, i.e., node 5 and node 6.
  }
  \label{fig_bkpivot}
\end{figure}

The pseudo-code is given in Alg. \ref{alg_pivot} in the Appendix.
The recursive procedure, BKPivot$(G,k,P,C,X)$, lists all maximal $k$-plexes that must subsume $P$, possibly include vertices in $C$ and must not contain any vertex in $X$.
The idea relies on the fact that, if $G[P\cup C]$ is a $k$-plex, then no further branches will be produced. Otherwise, there is a vertex in $P\cup C$ that has at least $k+1$ non-neighbors in $G[P\cup C]$, including itself.
In detail, BKPivot first checks the maximality of $P$.
Afterwards, a vertex $u_p$ of minimum degree in $G[P\cup C]$ is selected as pivot and BKPivot branches as follows:

\begin{itemize}
  \item If $u_p$ is not adjacent to at most $k$ vertices in $P\cup C$, then $G[P\cup C]$ is a $k$-plex. In this case, we check if $G[P\cup C]$ is maximal in $G$. If so, emit $P\cup C$ and stop the current branch.
  %by validating if vertex of $E$ can be added to $G[P\cup C]$ and forms a larger $k$-plex.
  \item Otherwise, $u_p$ is not adjacent to $q$ vertices in $P\cup C$, where $q\geq k+1$.
    The consecutive branches are generated with respect to either $u_p\notin P$ and $u_p\in P$.
    \begin{itemize}
    \item If $u_p \notin P$, we generate two branches by either moving $u_p$ from $C$ to $X$ or moving $u_p$ from $C$ to $P$. The latter case will fall into the next case.
    \item If $u_p\in P$, let $|P\setminus N(u_p)|=q_1$ and $|C\setminus N(u_p)|=q_2$. Then $q_1+q_2=q$. It is not hard to prove $q_1<k$ and let $k'=k-q_1$. Thus, at most $k'$ vertices in $C\setminus N(u_p)$ can be included in the $k$-plex.
  Denote $C\setminus N(u_p)$ as $\{u_1\cdots u_{q_2}\}$ by an arbitrary order. we generate $k'+1$ branches:
             \begin{itemize}
            \item[(a)] In the first branch, $u_1$ is moved from $C$ to $X$;
            \item[(b)] In the second branch, $u_1$ is moved from $C$ to $P$ and $u_2$ is moved from $C$ to $X$;
            \item[(c)] In the $i$th branch where $i$ is from 3 to $k'$, $\{u_1,...,u_{i-1}\}$ are moved from $C$ to $P$, and $u_i$ is moved from $C$ to $X$.
            \item[(d)] In the last branch, $\{u_1,...,u_{k'}\}$ are moved from $C$ to $P$ and $\{u_{k'+1},...,u_{q_2}\}$ are moved from $C$ to $X$.
          \end{itemize}
    \end{itemize}
\end{itemize}
Correctness relies on Steps (a)-(d). Each maximal $k$-plex will fall into one case of (a)-(d). In the last case (d), the maximal $k$-plexes that include $\{u_1,...,u_{k'}\}$ are visited. Because $u_p$ and $\{u_1,...,u_{k'}\}$ are in $P$, so $\{u_{k'+1},...,u_{q_2}\}$ can be excluded from further consideration since at most $k'$ vertices in $C\setminus N(u_p)$ can be included in the $k$-plex. \ Fig. \ref{fig_bkpivot}  shows an example of the branch scheme.

\subsection{Complexity analysis}
The main complexity result is below. See Appendix for the proof.

\begin{theorem}\label{Thm:R-time}
Given a graph $G=(V,E)$ with maximum degree $\Delta$ and degeneracy $D$, \emph{ListPlex}($G,k$) lists all maximal $k$-plexes without repetition in time $O(n^{2k}+n(D\Delta)^{k+1}\gamma_k^D)$, where $\gamma_k<2$ is the largest root of $1=x^{-1}+\dots+x^{-k-1}$. %\qed
\end{theorem}

\noindent
{\bf Remark}. Note that for each  $k$, the exponential part of the running time of our algorithm is $\gamma_k^D$ and $\gamma_k$ is bounded by $O(2- \frac{1}{2^{k+1}})$.
The exponential part for the Bron-Kerbosch algorithm is $2^{n}$. The exponential part for \citeauthor{conte2018d2k}'s decomposition algorithm is $2^{D\Delta}$,  The exponential part for Zhou et al.'s algorithm is $\gamma_k^{D\Delta}$. Hence, our algorithm provides a significant improvement of the previously known state-of-the-art algorithms.
By keeping the status of at most $k$ vertices at each branch, the \emph{BKPivot} also greatly optimizes the space complexity of Zhou et al.'s pivot heuristic.

\section{Listing Large Maximal $k$-Plexes}
In order to list large maximal $k$-plexes, i.e., maximal $k$-plexes of size at least $l$ ($l\ge 2k-1$), ListPlex can be reused by simply prohibiting the output of $k$-plexes smaller than $l$.
However, this mildly changed algorithm is previewed to be inefficient in practice.
In fact, it is possible to prune some branches early and improve the practical performance due to the import of this size constraint.
For example, because $l\ge 2k-1$, the search for maximal $k$-plexes of size at most $2k-2$ (Part I of Alg \ref{alg_ListPlex}) can be simply dropped.
For more stronger pruning techniques, let us first introduce an important observation.

\begin{lemma}
\label{lemma_size}
Assume $G[P]$ is a $k$-plex of $G=(V,E)$, $|P|\ge l$. %Denote $N_{P}(u,v)=N(u)\cap N(v) \cap P$.
For any two vertices $u,v \in P$, if $(u,v)\in E$, then $|N(u)\cap N(v) \cap P| \ge l-2k$, otherwise $|N(u)\cap N(v) \cap P| \ge l-2k+2$.
\end{lemma}

Note that this property was also observed in \cite{conte2018d2k,zhou2020enumerating}.

\subsection{Pruning seed graph $G_i$}

Suppose the degeneracy ordering of $G=(V,E)$ is $\eta=v_1,...,v_n$.
Recall that when we search the maximal $v_i$-leaded  $k$-plexes, we build a seed graph $G_i$ which is an induced graph of $\{v_i\}\cup N_{\succ_\eta}(v_i)\cup N^2_{\succ_\eta}(v_i)$.
Denote the vertex and edge sets of $G_i$ are $V_i$ and $E_i$, respectively.
We show rules to reduce the scale of $G_i$.
%Assume $P$ is a \emph{seed set} composing $v_i$.
%The prune rules are given as follows.
%Based on Lemma \ref{lemma_size}, in Line 5 of Alg. \ref{alg_ListPlex}, $C_1$ and $C_2$ can be further pruned.
\begin{prune}
\label{prune_size}
%Let $v_i$ be a seed vertex and $G_i$ is the seed graph.
%Denote $N_{P}(u,v_i)=N_G(u)\cap N_G(v_i) \cap P$.
%Then any vertex $u$ which
Assume $u\in V_i$, if $u$ satisfies
\begin{itemize}
  \item $u\in N_{\succ_\eta}(v_i)$ and $|N(u)\cap N(v_i)\cap V_i| < l-2k$,
  \item or $u\in N_{\succ_\eta}^2(v_i)$ and $|N(u) \cap N(v_i)\cap V_i| < l-2k+2$,
\end{itemize}
\end{prune}
then $u$ can be excluded from $G_i$ without affecting the correctness of ListPlex.

%A further study of Prune Rule \ref{prune_size} discloses that the size of pruned $N_{\succ_\eta}(v_i)$ is bounded by $O(\frac{D}{q-2k+2})$ and the size of pruned $N_{\succ_\eta}^2(v_i)$ is bounded by $O(\frac{D\Delta}{q-2k+2})$. Therefore, $|V_i|$ is bounded by $O(\frac{D\Delta}{q-2k+2})$.

\subsection{Excluding unfruitful seed sets}

%In Alg. \ref{alg_ListPlex}, we build $\sum_{j=1}^{k-1}\binom{|N^2_{\succ_\eta}(v_i)|}{j}$ seed sets $P_s$ for listing $v_i$-leaded maximal $k$-plexes.
%Each seed set $P_s$ includes $v_i$ and a $k$-1 size-bounded subset of $N^2_{\succ_\eta}(v_i)$.
%Accompanied with each $P_s$, the seed candidate set $C_s=N_{\succ_\eta}(v_i)$ and seed exclusive set $X_s=N^2_{\succ_\eta}(v_i)\setminus S$ are also built.
%These three sets are given to the recursive listing algorithm to traverse all $k$-plexes.
Intuitively, if we can identify some unfruitful seed sets $P_s$, i.e., sets that are impossible to be a part of large $k$-plexes, we can save the forthcoming exponential search in $G_i$.
With this in mind, we make use of the following pruning rule.

\begin{prune}
\label{prune_common}
Given a seed graph $G_i=(V_i,E_i)$, a seed set $P_s=\{v_i\}\cup S$ where $S\subseteq N^2_{\succ_\eta}(v_i)$ and $|S|\le k-1$. Denote $C_s = N_{\succ_\eta}(v_i)$. For any two vertices $u,v \in S$, if
\begin{itemize}
  \item $(u,v)\in E$ and $|N_{G_i}(u) \cap N_{G_i}(v) \cap C_s| < l-2k-max(k-3,0)$,
  \item or $(u,v)\notin E$ and $|N_{G_i}(u) \cap N_{G_i}(v) \cap C_s| < l-2k+2-max(k-3,0)$.
\end{itemize}
then $P_s$ is not in any maximal $v_i$-leaded $k$-plexes of size at least $l$.
\end{prune}

\begin{figure}[htb]
  \centering
  % Requires \usepackage{graphicx}
  \includegraphics[width=1\columnwidth]{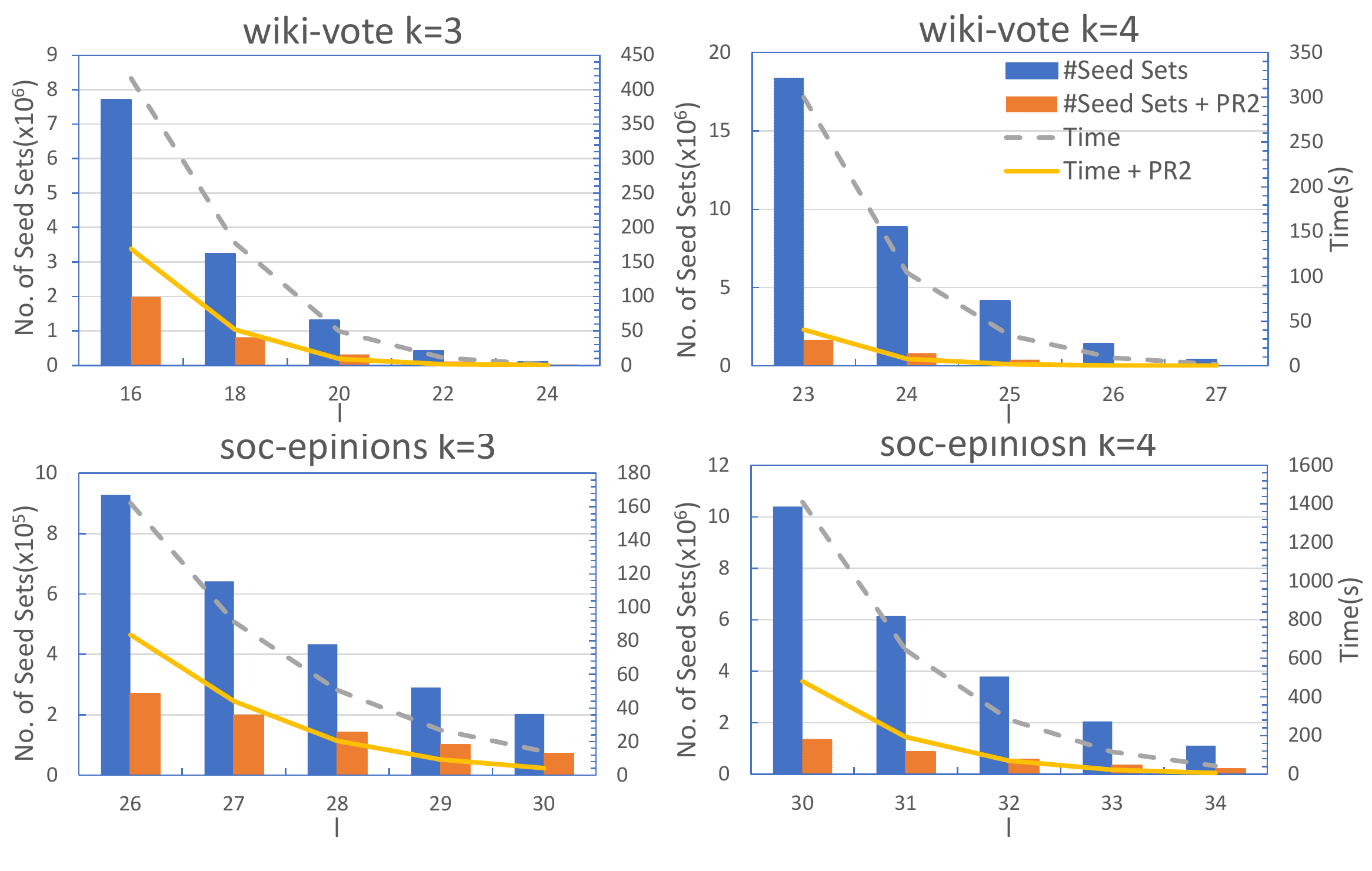}\\
  \caption{The number of seed sets and running time with and without Prune Rule \ref{prune_common}.}
\label{fig_prune2}
\end{figure}

It turns out that this prune rule dramatically improves the performance of our algorithm. In Fig. \ref{fig_prune2}, we show the comparison between the algorithm using Prune Rule \ref{prune_common} and the one without it.

\section{Implementation Techniques}
We present  important techniques to implement ListPlex on modern computers: computers with multi-level caches and multiple cores.

\subsection{Reducing cache misses}
\label{subsection_io_opt}
%In the implementation, the graph $G=(V,E)$ and the seed graph $G_i$ are kept by the Compressed Sparse Row (CSR) format \cite{chang2019cohesive}.
In an initial implementation, the algorithm searches maximal $k$-plexes by visiting $G_i$ and $G$ alternatively.
When a maximal $v_i$-leaded $k$-plex $G[P]$ is found from $G_i$, ListPlex revisits the input graph $G$ to validate if a
vertex in $v_1,...,v_{i-1}$ forms a larger $k$-plex with that solution.
This % initial implementation suffers
results in a high amount of cache misses when checking the maximality of $G[P]$. Clearly, it is partially caused by the fact that the data of $G$ is swapped out from the cache.

\begin{figure}[htb]
  \centering
  \includegraphics[width=1.0\columnwidth]{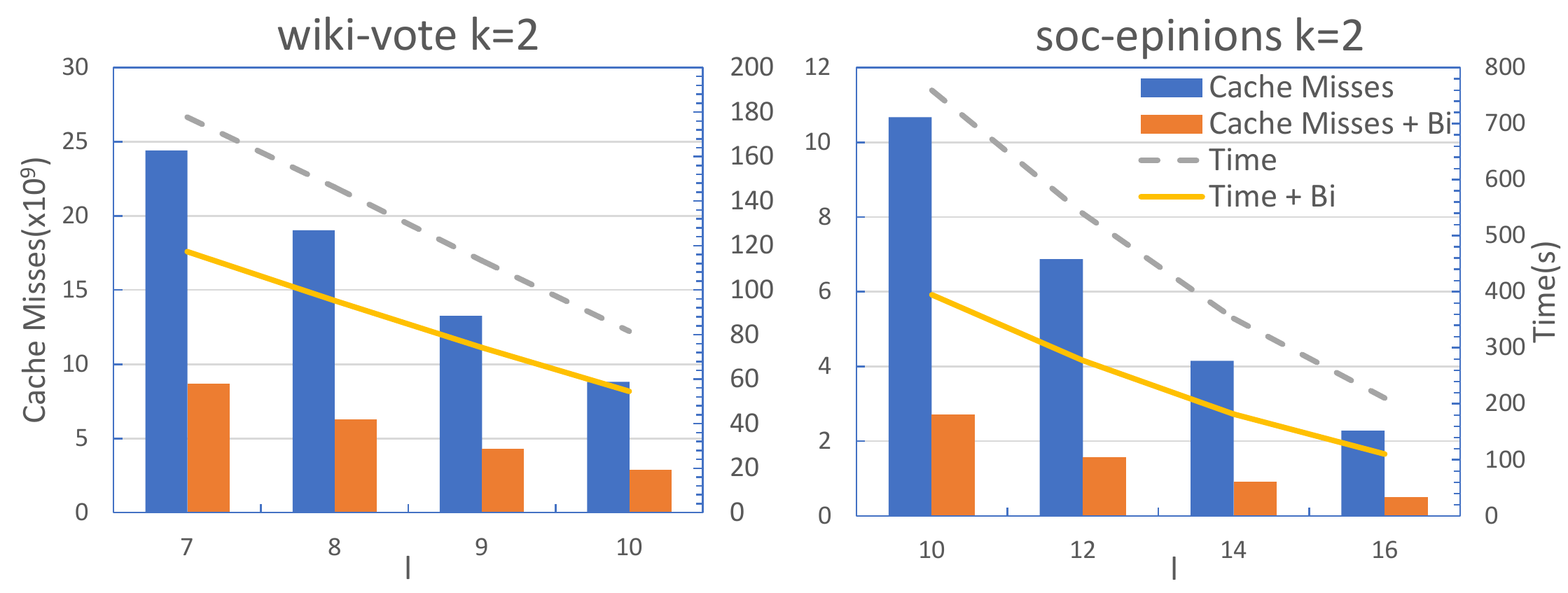}\\
  \caption{The total number of data cache misses and the running time with and without using bipartite graph $B_i$.}
  \label{fig_cache_example}
\end{figure}

In order to reduce cache misses, we further make use of the diameter-2 property of large $k$-plexes.
For each vertex $v_i$ in ordering $\eta$, we build a bipartite graph $B_i=(L_i, R_i, F_i)$ where $L_i=N_{\prec_\eta}(v_i) \cup N^2_{\prec_\eta}(v_i)$, $R_i=\{v_i\}\cup N_{\succ_\eta}(v_i) \cup N^2_{\succ_\eta}(v_i)$ and edge set $F_i\subseteq L_i\times R_i$ is induced from $G$.
When BKPivot finds a maximal $v_i$-leaded $k$-plex $G[P]$ on $G_i$, we further validate if for each vertex $u\in L_i$,
\begin{itemize}
  \item $|N_{B_i}(u) \cap P| \le |P|+1-k$ or
  \item $\exists v\in P$ that $|N_{G_i}(v)\cap P|=|P|-k$ and $(u,v) \notin E$.
\end{itemize}
then $G[P]$ is maximal in $G$.
With $B_i$, to check the maximality of a $k$-plex, we only need to visit $G_i$ and $B_i$. Though the vertex numbers of $G_i$ and $B_i$ are both $O(D\Delta)$, in real-world graphs, the vertex numbers of $G_i$ and $B_i$ are far less than $|V|$, implying good locality.
We compare the time and cache misses between the algorithm using $B_i$ and the one without $B_i$ in Figure \ref{fig_cache_example}.
%Note that $B_i$ is also stored with CSR format.

\vspace{-2mm}

\subsection{Parallelization}
ListPlex also owns appealing parallel features. We introduce a shared-memory parallel version of ListPlex in this subsection.

It is observed that searches of maximal $v_i$-leaded $k$-plexes are independent for each $v_i$.
Thus, for each vertex $v_i$, we create a task, say $T_i$, to process the search of all maximal $v_i$-leaded $k$-plexes.
$T_i$ owns its private seed graph $G_i$ and bipartite graph $B_i$. Tasks $T_1,...,T_n$ can be executed in parallel.

However, it could happen that most tasks stop but a few heavy tasks are still running.
Specifically, when the number of running tasks is less than the number of available cores, computational resources are wasted.
In such case, we split the branches of a running task $T_i$ into new subtasks for the idle cores.
Assume that $T_n$ has been dispatched for execution but $T_i$ ($i<n$) is still staying in the BKPivot$(G_i, k, P, C, X)$ procedure.
Then, when $T_i$ detects some idle cores, it spawns recursive calls to BKPivot$(G_i, k, P, C, X)$ as subtasks of $T_i$ and dispatches them to idle cores.
A subtask of $T_i$ owns its sets $P$, $C$ and $X$ but shares $G_i$ and $B_i$ with $T_i$.
Indeed, the schedule follows the work-stealing scheduling algorithm which accommodates well with the Bron-Kerbosch algorithm \cite{blanuvsa2020manycore}.

\paragraph{Fine-Grained Task} In parallel computing, the granularity of subtasks substantially affects the performance.
Empirically, simple small tasks should not be spawned due to the overhead.
In our implementation, we measure the complexity of a subtask, i.e., the time of executing BKPivot$(G_i, k, P, C, X)$,  by the size of $C$.
Particularly, if $|C|>10$ and there are some idle cores, we spawn new subtasks and assign them to available cores.

\section{Experiments}
%In this section, we carry out experiments in real-world graphs to evaluate the proposed algorithms $ListPlex$ and $ParaPlex$ for both Problem \ref{problem_diameter2} and Problem \ref{problem_q} (Community Detection).

\paragraph{Experiments setup}
The codes are written in C++11 and compiled by g++-9.3.0 with optimization option '-O3'. All experiments are conducted on a computer with a Ubuntu20.04 operating system, two-way Intel Xeon Gold 6130 CPUs (2.1GHz, 22MB L3-cache, 2 CPU chips and 32 physical cores in total), a 132G RAM and a 1T SSD.
%In order to reduce the uncertainty of our test environment,
We also disable hyper-threading and turbo techniques. ListPlex is parallelized with the OpenMP library.

%As for the more practical problems of listing large maximal $k$-plexes, i.e., listing maximal $k$-plexes of size at least $l$ where $l$ must be at least $2k-1$, the number of solvers becomes richer \cite{conte2018d2k,zhou2020enumerating}.
%As far as we know, D2K \cite{conte2018d2k} and CommuPlex \cite{zhou2020enumerating} are among the most recent and competitive solvers \footnote{We appreciate the authors of D2K \cite{conte2018d2k} and CommuPlex \cite{zhou2020enumerating} for providing their codes.}. Both D2K and CommuPlex are compiled with their makefiles.

% Since GP and LP are algorithms of enumerating maximal $k$-plexes, we compare \emph{FaPlexen} with LP and GP.
%Note that GP is an implementation of Bron-Kerbosch framework (i.e., \emph{Plexen}) which outperforms the previous implementation of \cite{wu2007parallel},  D2K is dedicated to find constrained maximal $k$-plex defined in Problem \ref{prob-com-detect1}. Hence, we compare \emph{FaPlexen} with LP and GP, then compare CommuPlex with D2K in massive real-life graphs.
\paragraph{Dataset}
In Table \ref{tbl-data}, we report basic information of benchmark graphs, including the number of vertices $n$, number of undirected edges $m$, maximum degree $\Delta$ and degeneracy $D$.
 These graphs are taken from Stanford Large Network Dataset Collection (SNAP) \cite{snapnets} and Laboratory for Web Algorithmics (LAW) \footnote{http://law.di.unimi.it/}.
As we can see, the size of these graphs broadly ranges. Like \cite{conte2018d2k}, we divide them into three categories, i.e., small, medium and large graphs.
Large graphs have more than ten million nodes, medium graphs are those with more than ten thousand nodes while the remaining graphs are classified as small graphs.
\begin{table}[h]
% \scriptsize
\centering
  \caption{Considered networks and their properties}
  \label{tbl-data}
  \resizebox{0.8\columnwidth}{!}{
      \begin{tabular}{c|c|c|c|c}
        \toprule[2pt]
        Network & n & m & $\Delta$ & D \\
        \hline
        jazz & 198 & 2742 & 100 & 29 \\
        ca-grqc & 5241 & 14484 & 81 & 43\\
        gnutella08 & 6301 & 41554 & 97 & 10\\
        wiki-vote & 7116 & 100763 & 1065 & 53\\
        lastfm & 7624 & 55612 & 216 & 20\\
        \hline
        as-caida & 26475 & 53381 & 2628 & 22\\
        soc-epinions & 75888 & 405739 & 3044 & 67\\
        soc-slashdot & 82144 & 500480 & 2548 & 54\\
        email-euall & 265214 & 365569 & 7636 & 37\\
        amazon0505 & 410236 & 2439436 & 2760 & 10\\
        in-2004 & 1353703 & 13126172 & 21869 & 488\\
        soc-pokec & 1632803 & 22301964 & 14854 & 47\\
        as-skitter & 1696415 & 11095298 & 35455 & 111\\
        soc-livejournal & 4847571 & 68993773 & 14815 & 360\\
        \hline
        arabic-2005 & 22744080 & 639999458 & 575628 & 3247\\
        uk-2005 & 39459925 & 936364282 & 1372171 & 584\\
        it-2004 & 41291594 & 1150725436 & 1243927 & 3209\\
        webbase-2001 & 118142155 & 1019903190 & 816127 & 1506\\
		\bottomrule[2pt]
\end{tabular}
}
\end{table}

\subsection{Listing all maximal $k$-plexes}

In this section, we evaluate the performance of our ListPlex for listing all maximal $k$-plexes.
We compare our ListPlex with the fastest known algorithm BKPivot \cite{zhou2020enumerating} and the traditional Bron-Kerbosch algorithm BKPlex. Note that the competitive D2K \cite{conte2018d2k} solver only outputs large maximal $k$-plexes, i.e., $k$-plexes of size at least $l$ where $l>2k-2$.
The recent solvers \emph{GP} \cite{wang2017parallelizing} and \emph{Enum} \cite{berlowitz2015efficient} are not as time-efficient as BKPivot, see  \cite{zhou2020enumerating}. 
%\textcolor{green}{(GP and Enum are implemented in Java and Python, respectively).[May be not important?]}
In case a solver cannot finish in 12 hours (43200 seconds) for an instance, we imperatively stop it. In the table, we mark the unfinished instances with OOT.

In Table \ref{tbl-pro1}, we show the time performance of these listing algorithms.
We also report the parallel running time of ListPlex with 16 threads and the parallel speedup.
Due to the huge amount of maximal $k$-plexes, neither of these algorithms is  able to list all of them on medium or large graphs in 12 hours, even setting $k=2$.

In terms of time, ListPlex outperforms both competitors for all these instances.
For cases like wiki-vote with $k=2$, ListPlex runs like 7$\times$ faster than the other algorithms.
ListPlex also achieves a nearly perfect speedup for almost all cases except very simple ones.
Unexpectedly, BKPlex runs faster than BKPivot for the last two larger graphs when $k=2$. %implying that we may alter to BKPlex for different seed graphs in our algorithm.
%However, it is out of scope to discuss when BKPLex or BKPivot is preferred.

\begin{table}[h]
\centering
  \caption{Listing all maximal $k$-plexes in small graphs}
  \label{tbl-pro1}
   \resizebox{1\columnwidth}{!}{
    \begin{tabular}{c|c|c|c|c|c|c|c}
    \toprule[2pt]
    \multirow{2}{*}{Network} & \multirow{2}{*}{$k$} & \multirow{2}{*}{\#$k$-plexes} & \multicolumn{4}{c|}{The running time (s)} & \multirow{2}{*}{Speedup}\\
	\cline{4-7}
	& & & BKPlex & BKPivot & ListPlex & ListPlex(16)&\\
	\hline
    jazz  & 2     & 35214 & 648.864 & 0.29 & \textbf{0.086} & 0.408 & 0.211 \\
    jazz  & 3     & 3602575 & 772.826 & 17.55 & \textbf{6.477} & 0.832 & 7.785 \\
    jazz  & 4     & 193056583 & 3226.746 & 829.40 & \textbf{417.646} & 26.187 & 15.949 \\
    \hline
    ca-grqc & 2 & 13718439 & OOT & 1858.02 & \textbf{649.985} & 40.880 & 15.899\\
    \hline
    gnutella08 & 2 & 19866959 & 1500.208 & 3627.57 & \textbf{1117.858} & 70.207 & 15.922\\
    \hline
    wiki-vote & 2 & 66193264 & 10356.553 & 10671.92 & \textbf{1526.884} & 95.656 & 15.962\\
    \hline
    lastfm & 2 & 29086855 & 2643.394 & 6676.89 & \textbf{1989.701} & 124.525 & 15.978\\
    \bottomrule[2pt]
    \end{tabular}
    }
\end{table}

%  Besides the graph size, the number of $k$-plexes also increases exponentially with $k$ \cite{conte2018d2k}.

% Table generated by Excel2LaTeX from sheet 'Sheet1'
\begin{table*}[t]
    \centering
	\caption{The running time of listing large maximal $k$-plexes from small and medium graphs by CommuPlex, D2K and ListPlex.}
	\label{tbl-community1}
    \resizebox{1.8\columnwidth}{!}{
    \begin{tabular}{c|c|c|c|c|c|c|c|c|c|c|c|c|c}
	\toprule[2pt]
	\multirow{2}{*}{\tabincell{c}{Graph\\ $(|V|,|E|)$}} & \multirow{2}{*}{$k$}    & \multirow{2}{*}{$l$} & \multirow{2}{*}{\#$k$-plexes}  &  \multicolumn{3}{c|}{The running time (s)}   & \multirow{2}{*}{\tabincell{c}{Graph\\ $(|V|,|E|)$}} & \multirow{2}{*}{$k$}    & \multirow{2}{*}{$l$} & \multirow{2}{*}{\#$k$-plexes}  &  \multicolumn{3}{c}{The running time (s)}
	\\
	\cline{5-7} \cline{12-14}
	& & & & CommuPlex & D2K & ListPlex & & & & & CommuPlex & D2K & ListPlex\\
	\hline
	
	\tabincell{c}{jazz (198, 2742)}
	& 4 & 12 & 2745953 & 25.218 & 33.054 & \textbf{4.498} &\multirow{9}{*}{\tabincell{c}{wiki-vote\\ (7116, 100763)}} & \multirow{3}{*}{2} & 12 & 2919931 & 75.871 & 115.757  & \textbf{17.653}  \\
	\cline{1-7}
    \tabincell{c}{lastfm (7624, 55612)} & 4 & 12 & 1827337 & 20.724 & 23.991 & \textbf{4.586} &  &  & 20 & 52 & 4.52 & 11.289 & \textbf{0.591} \\
    \cline{1-7}
    \multirow{2}{*}{\tabincell{c}{as-caida\\(26475, 53381)}} & 3 & 12 & 281251 & 5.684 & 13.421 & \textbf{0.867} & & & 30 & 0 & 1.033 & \textbf{0.027} & 0.091 \\
    \cline{2-7}  \cline{9-14}
     & 4 & 12 & 15939891 & 300.388 & 785.506 & \textbf{47.98} & & \multirow{3}{*}{3} & 12 & 458153397 & OOT & OOT & \textbf{2185.598} \\
    \cline{1-7}
    \multirow{3}{*}{\tabincell{c}{amazon0505\\(410236, 2439436)}} & 2 & 12 & 376 & 1.825 & 0.641 & \textbf{0.137} & & & 20 & 156727 & 595.636 & 1852.186 & \textbf{9.384} \\
    \cline{2-7}  & 3 & 12 & 6347 & 11.359 & 0.77  & \textbf{0.286} & & & 30 & 0 & 1.072 & \textbf{0.029} & 0.1 \\
    \cline{2-7} \cline{9-14} & 4 & 12 & 105649 & 47.049 & 5.338 & \textbf{1.171} & & \multirow{2}{*}{4} & 20 & 46729532 & OOT & OOT & \textbf{1174.2} \\
    \cline{1-7}   \multirow{4}{*}{\tabincell{c}{email-euall\\(265214, 365569)}} & 2 & 12 & 412779 & 8.793 & 11.199 & \textbf{1.946} & & & 30 & 0 & 9.17 & 3.627 & \textbf{0.112} \\
    \cline{2-14} & \multirow{2}{*}{3} & 12 & 32639016 & 619.384 & 1043.266 & \textbf{91.62} & \multirow{8}{*}{\tabincell{c}{soc-pokec\\(1632803, 22301964)}} & \multirow{3}{*}{2} & 12 & 7679906 & 1537.506 & 172.987 & \textbf{47.475} \\
    & & 20 & 2637 & 10.754 & 53.691 & \textbf{0.429} & & & 20 & 94184 & 1064.371 & 20.03 & \textbf{15.161} \\
    \cline{2-7} & 4 & 20 & 1707177 & 825.126 & 3800.889 & \textbf{24.089} & & & 30 & 3 & 662.64 & \textbf{8.637} & 9.557 \\
    \cline{1-7} \cline{9-14} \multirow{7}{*}{\tabincell{c}{soc-slashdot\\(82144, 500480)}} & \multirow{3}{*}{2} & 12 & 27208777 & 376.071 & 213.141 & \textbf{59.42} & & \multirow{3}{*}{3} & 12 & 520888893 & OOT & OOT & \textbf{1607.285} \\
    & & 20 & 11411028 & 227.016 & 137.159 & \textbf{32.988} & & & 20 & 5911456 & 1470.536 & 856.393 & \textbf{46.262} \\
    & & 30 & 453 & 10.77 & 16.481 & \textbf{0.688} & & & 30 & 5 & 717.425 & \textbf{9.993} & 10.127 \\
    \cline{2-7} \cline{9-14} & \multirow{3}{*}{3} & 12 & 2807943240 & OOT & 26029.006 & \textbf{7813.045} & & \multirow{2}{*}{4} & 20 & 318035938 & 34048.155 & OOT & \textbf{1825.216} \\
    & & 20 & 1303148522 & 28361.707 & 15308.777 & \textbf{4538.022} & & & 30 & 4515 & 1140.117 & 111.987 & \textbf{11.211} \\
    \cline{8-14} & & 30 & 1679468 & 699.876 & 2066.598 & \textbf{51.364} & \multirow{6}{*}{\tabincell{c}{soc-epinions\\(75888, 405739)}} & \multirow{3}{*}{2} & 12 & 49823056 & 843.9 & 735.589 & \textbf{193.307} \\
    \cline{2-7} & 4 & 30 & 502699966 & OOT & OOT & \textbf{6680.261} & & & 20 & 3322167 & 137.427 & 180.061 & \textbf{19.382} \\
    \cline{1-7} \multirow{4}{*}{\tabincell{c}{as-skitter\\(1696415, 11095298)}} & 2 & 50 & 47969775 & OOT & OOT & \textbf{520.884} & & & 30 & 0 & 8.995 & 12.109 & \textbf{0.492} \\
    \cline{9-14} & 2 & 100 & 0 & 1.793 & 2.951 & \textbf{0.716} & & \multirow{2}{*}{3} & 20 & 548634119 & 27037.614 & 35525.693 & \textbf{3072.267} \\
    \cline{2-7} & 3 & 50 & 21070497438 & OOT & OOT & OOT & & & 30 & 16066 & 546.69 & 2591.439 & \textbf{6.123} \\
    \cline{9-14} & 3 & 100 & 0 & 2.37 & 3.285 & \textbf{0.718} & & 4 & 30 & 13172906 & OOT & OOT & \textbf{661.103} \\
    \cline{1-7} \cline{8-14} \multirow{4}{*}{\tabincell{c}{in-2004\\(1353703, 13126172)}} & 2 & 50 & 25855779 & 7663.843 & 576.06 & \textbf{150.212} &\multirow{4}{*}{\tabincell{c}{com-livejournal\\(4847571, 68993773)}} & 2 & 340 & 650322 & 2284.435 & OOT & \textbf{109.382} \\
    & 2 & 100 & 9978037 & 5899.638 & 256.225 & \textbf{72.063} & & 2 & 345 & 0 & 57.548 & 13589.487 & \textbf{6.914} \\
    \cline{2-7} \cline{9-14} & 3 & 50 & 29045783792 & OOT & OOT & OOT & & 3 & 340 & 555718694 & OOT & OOT & \textbf{22863.467} \\
    & 3 & 100 & 4257410159 & OOT & OOT & \textbf{28384.76} & & 3 & 345 & 3963139 & 24861.871 & OOT & \textbf{826.183} \\
    \bottomrule[2pt]
    \end{tabular}
    }
\end{table*}

% The $2^{nd}$ DIMACS graphs are well-known difficult benchmarks for clique problems.
% We then compare the running times of \emph{FaPlexen} with GP and LP on all 80 $2^{nd}$ DIMACS graphs with $k$=2, 3, 4 and 5.
% In Fig. \ref{real_data}, we show the number of solved graphs against the elapsed time.
% Clearly,  \emph{FaPlexen} solves more instances than GP and LP for all $k$s.
% LP can only solves 1 to 2 graphs when $k=3$ and $4$, possibly caused by the fact that it does not optimize the worst-case running time.
% As $k$ grows, the problem becomes harder since all the algorithms can solves fewer instances in the same time frame. We publish the detailed computational results of these graphs along with the codes.

%However, we observe an exception for MANN-a9 (${|E|}/{\binom{|V|}{2}}=0.92$).

%In Table \ref{tbl-real}, we compare  the 3 algorithms on 5 real graphs which were tested in \cite{wang2017parallelizing}. The graph caida is omitef cannot be solved by any of the algorithms in 1 day. Results of the other 4 graphs are presented in Table \ref{tbl-real}.
%An unique exception is observed for the $G(100,0.1)$ when $k=5$.
%In random and real graphs, the results generally show that the running time of \emph{FaPlexen} scales slower than GP when $k$ grows.

\subsection{Listing large maximal $k$-plexes}

We evaluate the problem of listing large maximal $k$-plexes, i.e., maximal $k$-plexes that have at least $l$ vertices.
There are a rich number of solvers, e.g., GP \cite{wang2017parallelizing}, LP \cite{conte2017fast}, D2K \cite{conte2018d2k} and CommuPlex \cite{zhou2020enumerating} for the problem.
According to their empirical results, D2K and CommuPlex outperform earlier GP and LP in terms of practical running time.
Thus, we compare our ListPlex with D2K and CommuPlex in this subsection.
Also, D2K only outputs diameter-2 bounded maximal $k$-plexes. By setting $l$ at least $2k-1$, we make sure that three compared algorithms output the same set of $k$-plexes.
Also, we set a cut-off time of 12 hours for each instance.

%Since D2K, CommuPlex and ListPlex deal with the same problem when size constraint $l > 2k-2$.
\paragraph{The Sequential Performance}
Let us first compare the sequential versions of D2K, CommuPlex and ListPlex.
In Table \ref{tbl-community1}, we show the sequential running time of different algorithms.
For small networks, we set $k=2, 3$ and $4$, and $l=12, 20$ and $30$.
%When $k$ increases to a larger value, the running time of these algorithms dramatically raises.
For medium networks, we also set $k=2, 3$ and $4$ but we change $l$ for different graphs, mainly because all three algorithms cannot list all the $2$ to $4$-plexes even $k=30$ in 12 hours.
As for the large networks, we leave the test in the parallel environment.
These large graphs contain a dramatic number of maximal $k$-plexes that cannot be efficiently listed by these sequential algorithms.

ListPlex is the best performing algorithm for these instances.
Exceptions can only be observed in graphs which contain very few maximal $k$-plexes, e.g., wiki-vote with $k=2$ and $l=30$.
For the rest of these instances, ListPlex achieves a 4-100$\times$ speedup over CommuPlex and a 3-420$\times$ speedup over D2K.
For example, ListPlex is able to list all $4$-plexes with $l=20$ for wiki-vote in half an hour but CommuPlex and D2K cannot finish in 12 hours.
For some instances like soc-slashdot with $k=4$ and $l=30$, ListPlex is the only algorithm that lists all maximal $k$-plexes of size at least $l$.
It is worth observing that, the running time of D2K and CommuPlex contrasts in different scenarios, e.g., D2K runs 10$\times$ faster than CommuPlex in in-2004 with $k=2$ but CommuPlex performs much better in soc-epinions with $k=2$ or $3$.
In total, the results show the great superiority of ListPlex over the existing algorithms.

\begin{figure}[htb]
  \centering
  \includegraphics[width=0.8\columnwidth]{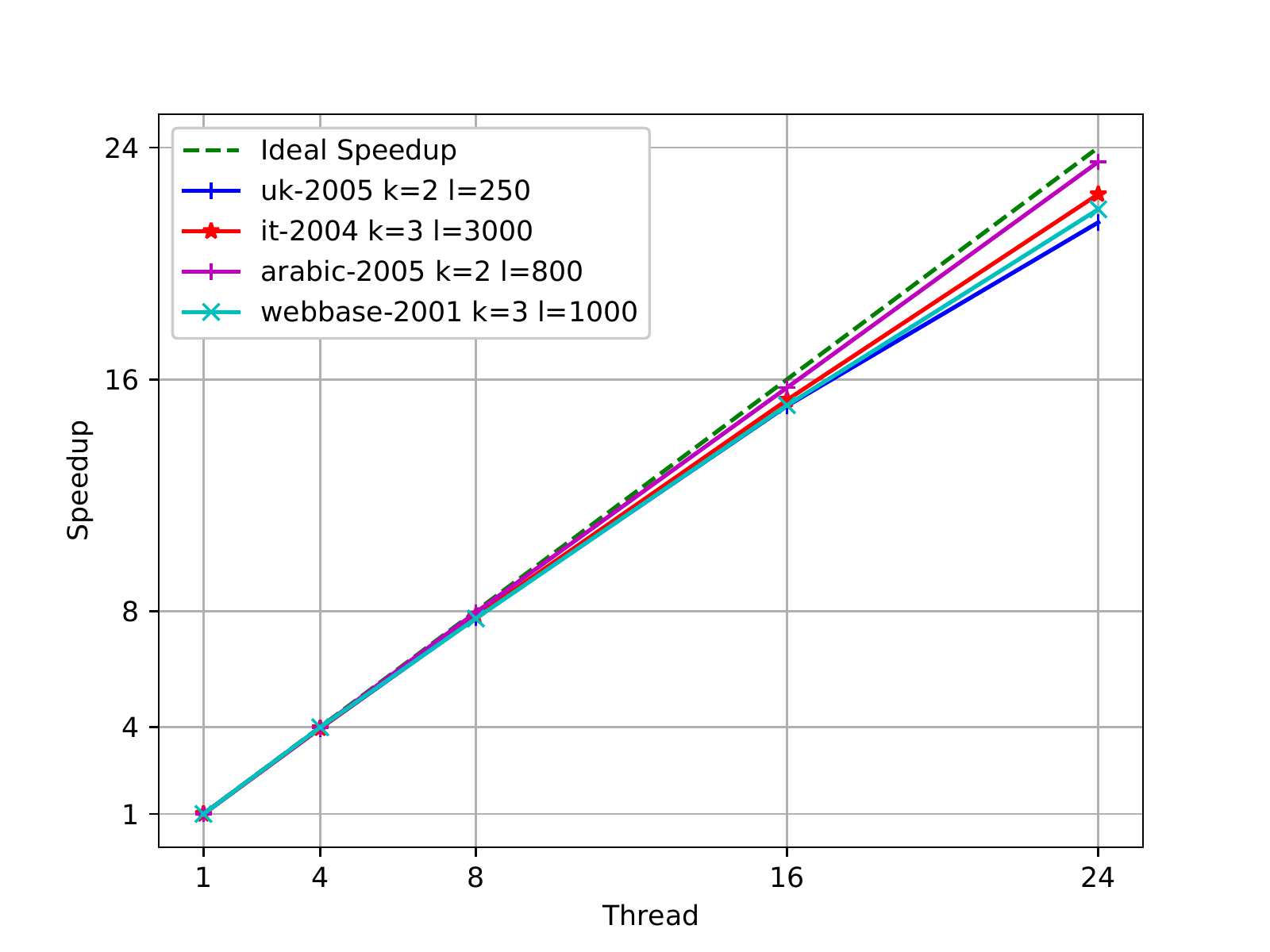}\\
  \caption{The speedup of ListPlex for the large graphs with different parameters. }
  \label{fig_speedup}
\end{figure}

\paragraph{The Parallel Performance}
%Note that both ListPlex and D2K have parallel versions.
It is known that D2K also provides a parallel version that achieves almost linear speedup for many instances.
In Table \ref{tbl-community2}, we run the parallel ListPlex and D2K with 16 threads for large networks.
Still, ListPlex runs about  3-8$\times$ faster than D2K in these tested instances.
In Fig. \ref{fig_speedup}, we show the speedup achieved by ListPlex for large graphs with different $k$s and $l$s.
Clearly, ListPlex also can reach a nearly perfect speedup in these instances.
As both ListPlex and D2K scale well in large graphs, the superiority of ListPlex may be achieved by doing fewer work.

% TODO
\begin{table}[t]
    \centering
	\caption{The parallel running time of large networks by ListPlex and D2K with 16 threads.}
	\label{tbl-community2}
    \resizebox{0.9\columnwidth}{!}{
    \begin{tabular}{c|c|c|c|c|c}
	\toprule[2pt]
	\multirow{2}{*}{\tabincell{c}{Graph\\ $(|V|,|E|)$}} & \multirow{2}{*}{$k$}    & \multirow{2}{*}{$l$} & \multirow{2}{*}{\#$k$-plexes}  &  \multicolumn{2}{c}{The running time (s)}
	\\
	\cline{5-6}
	& & & & D2K(16) & ListPlex(16)\\
    \hline \multirow{4}{*}{\tabincell{c}{arabic-2005\\(22744080, 639999458)}} & 2 & 800 & 224870903 & 2195.272 & \textbf{714.159}\\
    & 2 & 1000  & 236897 & 151.328 & \textbf{40.202}\\
    \cline{2-6}
    & 3 & 800 & $>$25062182205 & OOT & OOT\\
    & 3 & 1000 & 34155502 & 587.967 & \textbf{128.737}\\
	\hline \multirow{4}{*}{\tabincell{c}{uk-2005\\(39459925, 936364282)}} & 2 & 250 & 106243475 & OOT & \textbf{355.855}\\
    & 2 & 500 & 256406 & 318.118 & \textbf{35.001}\\
    \cline{2-6}
    & 3 & 250 & $>$18336111409 & OOT & OOT\\
    & 3 & 500 & 28199814 & 9506.661 & \textbf{121.726}\\
    \hline \multirow{4}{*}{\tabincell{c}{it-2004\\(41291594, 1150725436)}} & 2 & 2000 & 675111 & 340.904 & \textbf{41.983} \\
    & 2 & 3000 & 675111 & 307.735 & \textbf{38.468} \\
    \cline{2-6}
    & 3 & 2000 & 197679229 & 4254.456 & \textbf{724.979}\\
    & 3 & 3000  & 197679229 & 4235.389 & \textbf{715.002} \\
    \hline \multirow{4}{*}{\tabincell{c}{webbase-2001\\(118142155,  1019903190)}} & 2 & 800 & 1599005 & 374.134 & \textbf{54.19} \\
    & 2 & 1000 & 1164383 & 346.393 & \textbf{53.651} \\
    \cline{2-6}
    & 3 & 800 & 1785341050 & 36116.817 &\textbf{5521.386} \\
    & 3 & 1000 & 1484341137 & 35005.343 & \textbf{6960.816} \\
    \bottomrule[2pt]
    \end{tabular}
    }
\end{table}

% For the computation of4-plexes with��=30for graphs Slashdot090221 and soc-Epinions1,CommuPlexwasas able to found the all communities whileD2Knot. For graphs likesoc-Epinions1, jazz and CA-GrQc, our algorithm outperformsD2Kfor all��s and��s, while for the other graphs,

\section{Conclusion}
We studied the problems of listing maximal $k$-plexes and maximal $k$-plexes of prescribed size.
We proposed ListPlex, a fast and scalable algorithm that efficiently solves the two problems in real-world graphs.
Especially, ListPlex combines a new decomposition scheme with the branching algorithm, achieving a better theoretical complexity.
When maximal $k$-plexes of size at least $l$ ($l\ge 2k-1$) are asked, ListPLex can be also used for listing these large maximal $k$-plexes.
For practical considerations, we designed some additional prune rules for listing large maximal $k$-plexes.
These prune rules work very well in the context of large real-world graphs.
Furthermore, we designed a new local bipartite graph to improve the cache performance of the algorithm, and parallel scheduling strategies to increase parallelism.
Extensive empirical evaluations show the superiority of ListPlex over the state-of-the-art approaches for both problems.

%\textcolor{green}{
%Our work not only provides new insights to the fundamental problem of listing maximal k-plexes but also paves the road of the utilization k-plex model in future web mining tasks.
%}
% In parallel scenario, we parallelized ListPlex and obtained a nearly linear speedup.
%Both theoretical analysis and extensive empirical evaluation show that the proposed algorithms have great superiority over the state-of-the-art algorithms in terms of both worst-case time complexity and computation experiments.

% Our work not only provides new insights to the fundamental problem of listing maximal $k$-plexes, but also paves the road of the utilization $k$-plex algorithm in many graph mining tasks.

%%
%% The acknowledgments section is defined using the "acks" environment
%% (and NOT an unnumbered section). This ensures the proper
%% identification of the section in the article metadata, and the
%% consistent spelling of the heading.
\begin{acks}
This work is supported by National Natural Science Foundation of China under grant nos. 61802049, 61972070 and 62172077.
\end{acks}

%%
%% The next two lines define the bibliography style to be used, and
%% the bibliography file.
\newpage
\bibliographystyle{ACM-Reference-Format}
\bibliography{main}

\newpage
\appendix
\section{The Bron-Kerbosch Algorithm}

\begin{algorithm}[ht!]
    \DontPrintSemicolon
    % \scriptsize
        \caption{The Basic Bron-Kerbosch Algorithm }
        \label{alg_BK}
        % \KwIn{a graph $G=(V,E)$, an integer $k>1$}
        % \KwOut{all vertex sets that induce maximal $k$-plexes in $G$. }
        \emph{BKPlex}$(G,k)$\\
        \Begin{
            \emph{BKRec}$(G,k,\emptyset, V,\emptyset)$
        }
        \emph{BKRec}$(G,k, P, C, X)$\\
        \Begin{
              $C\gets\{v\in C: G[P\cup\{v\}]\mbox{ is a $k$-plex}\}$\\
              $X\gets\{v\in X: G[P\cup\{v\}]\mbox{ is a $k$-plex}\}$\\
           \If {$C= \emptyset$}{
                \If{$X= \emptyset$}{
                    emit $P$ \\
                }
               \Return
            }
            \Else{
                \For{ $u\in C$}{
                    $C\gets C \setminus \{u\}$\\
                    %$Cand'\gets\{v\in Cand: Plex\cup\{v\}\mbox{ is a $k$-plex}\}$\\
                   %$Excl'\gets\{v\in Excl: Plex\cup\{v\}\mbox{ is a $k$-plex}\}$\\
                    \emph{BKRec}$(G, k, P\cup\{u\}, C, X)$ \\
                    $X\gets X\cup \{u\}$
                }
            }
        }
        %\emph{update}$(G, k, Plex, Cand, Excl)$\\
        %\Begin{
        %    $Cand'\gets\{v\in Cand: Plex\cup\{v\}\mbox{ is a $k$-plex}\}$\\
        %    $Excl'\gets\{v\in Excl: Plex\cup\{v\}\mbox{ is a $k$-plex}\}$\\
        %    \Return {Cand',Excl'}
        %}
\end{algorithm}

\section{The BKPivot algorithm}
\begin{algorithm}[ht!]
    \DontPrintSemicolon
    % \scriptsize
        \caption{The Bron-Kerbosch algorithm with pivot heuristic for listing all maximal $k$-plexes.}
        \label{alg_pivot}
       % \KwIn{a graph $G=(V,E)$, an integer $k>1$, disjoint sets $P,C$ and $X \subseteq V$. $G[P]$ must be a $k$-plex.}
      %  \KwOut{all vertex sets that induce maximal $k$-plexes. The vertex set must include $P$, possibly include some vertices $C$ and  must not include $X$ form $G$. }
        %\tcc{For brevity, we omit $G'_>,G,k,q$ in subcalls}
        \emph{BKPivot}$(G,k,P,C,X)$\\
        \Begin{
             $C \gets \{v: v\in C \mbox{ and $G[\{v\}\cup{P}]$ is a $k$-plex} \}$\\
             $X \gets \{v: v\in X \mbox{ and $G[\{v\}\cup{P}]$ is a $k$-plex} \}$ \\
             \If {$C= \emptyset$}{
                \If{$X= \emptyset$}{
                    emit $P$ \\
                }
               \Return
            }
            Find a vertex of minimum degree $u_p$ in $G[P\cup C]$\\
            \If{$|N(u_p)|\ge |P|+|C|-k$}{
                \If{$\nexists v\in X$ that $G[P\cup C \cup \{v\}]$ is a k-plex}{
                    emit $P\cup C$
                }
            }\ElseIf {$u_p\in P$}{
                    Let $u_1,...,u_{q_2}$ be an arbitrary ordering of $C\setminus N(u_p)$\\
                    $k'\gets k-|P\setminus N(u_p)|$\\
                    \emph{BKPivot}$(G,k,P,C\setminus \{u_1\},X\cup\{u_1\})$ \\
                     \For{$i\in \{2,...,k'\}$}{
                         \emph{BKPivot}$(G,k,P\cup \{u_1,...,u_{i-1}\}, C\setminus \{u_1,...,u_{i}\}, X\cup \{u_i\})$ \\
                     }
                    \emph{BKPivot}$(G,k,P\cup \{u_1,...,u_k'\}, C\setminus\{u_1,...,u_{q_2}\}, X)$\\

            }\Else{
                \emph{BKPivot}$(G,k,P,C\setminus \{u_p\},X\cup\{u_p\})$ \\
                \emph{BKPivot}$(G,k,P\cup\{u_p\}, C\setminus \{u_p\}, X)$ \\

            }
        }
\end{algorithm}

\section{Missing Proofs}
Proof of Lemma \ref{connectedP}
\begin{proof}
Let $u$ and $v$ be any pair of nonadjacent vertices in a $k$-plex. There are at most $k-1$ vertices not adjacent to $u$ and at most $k-1$ vertices not adjacent to $v$.
If the graph has more than $2k-2$ vertices, then there exists a vertex $w$ that is adjacent to both of $u$ and $v$. So the graph is connected and the diameter is at most 2.

Here is an example of a disconnected $k$-plex of size $2k-2$. The graph consists of two cliques of size $k-1$. We can see that the graph is a $k$-plex since each vertex is not adjacent to $k$ vertices (including itself).
The number of vertices in the graph is $2k-2$.
\end{proof}

\paragraph{Proof of Lemma \ref{lemma_diam2}}
\begin{proof}
By Lemma~\ref{connectedP}, we know that the diameter of $G[P]$ is bounded by 2. Since $G[P]$ contains $v_i$, we know that $G[P]$ can only be a subgraph of $G[\{v_i\}\cup N_{\succ_\eta}(v_i)\cup N^2_{\succ_\eta}(v_i)]$.
The second claim holds due to the definition of $k$-plexes.
\end{proof}

\paragraph{Proof of Lemma \ref{lemma_size}}
\begin{proof}
Let us denote $O=P\setminus \{u,v\}$. Then $|O|\ge q - 2$.
\begin{itemize}
    \item If $u$ and $v$ are adjacent, %then by definition of $k$-plex,
    there are at most $2(k-1)$ vertices that are not common neighbors of $u$ and $v$ in $O$. Thus, $|N(u)\cap N(v) \cap P|\ge |O|-2(k-1)\ge (l-2)-(2k-2)=l-2k$.
    \item If $u$ and $v$ are not adjacent, then there are at most $2(k-2)$ non-neighbors in $O$. Thus $|N(u)\cap N(v) \cap P|\ge |O|-2(k-2)\ge (l-2)-(2k-4)=l-2k+2$.
\end{itemize}
\end{proof}

\paragraph{Proof of Theorem \ref{Thm:R-time}}
\begin{proof}
For the first part, the running time is bounded by the number of subsets of size at most $2k-2$ times  the time to check its maximality.
There are at most $O({n\choose 2k-2})=O(n^{2k-2})$ subsets of size at most $2k-2$.
By \cite{zhou2020enumerating}, the time to check the maximality of a $k$-plex is bounded by $O(n^2)$.
So the running time is $O(n^{2k})$.

Before analyzing the part for listing maximal $k$-plexes of size at least $2k-1$, we first consider the running time bound of the procedure BKPivot. When $C=\emptyset$, we do not need to branch anymore. So we analyze our branching operations by measuring the number of vertices removed from $C$. The branching operation for the case $u_p\in P$ will generate $k'+1$ subbranches. In the first subbranch, one vertex $u_1$ is removed from $C$. In the second subbranch,
two vertices $\{u_1,u_2\}$ are removed from $C$. In the $i$th branch for $3\leq i \leq k'$, exactly $i$ vertices $\{u_1,\dots, u_{i}\}$ are removed from $C$.
In the last branch, $q_2$ vertices $\{u_{1},...,u_{q_2}\}$ are removed from $C$, where $q_2\geq k'+2$. If we use $T(c)$ to denote the running time of BKPivot working on $C$ with $c=|C|$,
then we get the following recurrence
\[ T(c)\leq T(c-1)+\dots +T(c-k')+T(c-q_2).
\]
When $u_p\notin P$ ($u_p\in C$), we generate two branches each of which will remove one vertex $u_p$ from $C$. In the latter case, we will follow with the above recurrence.
Combining them together, we have
 \[ T(c)\leq T(c-1)+\dots +T(c-k'-1)+T(c-q_2-1).
\]
Note that $k'\leq k-1$ and $q_2\geq k'+1$. For the worst case that $k'= k-1$ and $q_2= k'+1$, we get the recurrence
 \[ T(c)\leq T(c-1)+\dots +T(c-k)+T(c-k-1).
\]
Let $\gamma_k$ be the largest root of function $1=x^{-1}+\dots+x^{-k-1}$. Then the running time bound of the algorithm is bounded by $O(\gamma_k^{|C|})$. In our algorithm, initially $C$ is $N(v_i)$ and then $|C|\leq D$, where $D$ is the degeneracy of the graph.
We also note that $\gamma_k$ is strictly smaller than $2$. For example, when $k=1,2,3,4$ and $5$, $\gamma_k=1.618, 1.839,1.928,1.966$ and $1.984$, respectively.  Details on solving recurrence relations and time analysis can be found in \cite{exactbook}.

Next, we analyze the algorithm for listing maximal $k$-plexes of size at least $2k-1$. Note that computing the degeneracy order of a graph $G$ is in $O(m)$ %by a peeling algorithm
\cite{batagelj2003m}.  For each vertex $v_i$ in the degeneracy order, we find all maximal $v_i$-leaded $k$-plexes in the subgraph $G_i=G[\{v_i\}\cup N_{\succ_\eta}(v_i)\cup N^2_{\succ_\eta}(v_i)]$.
Hereby, we enumerate all subsets $S\subseteq N^2_{\succ_\eta}(v_i)$ with size $|S|\le k-1$ and for each $S$ we include it to $P$ to generate an instance.
So we will generate at most $|N^2_{\succ_\eta}(v_i)|^{k}$ instances. For each instance, we will call BKPivot with running time $O(\gamma_k^{|N_{\succ_\eta}(v_i)|})$.
Additionally, in order to validate the maximality of a maximal $v_i$-leaded $k$-plex in $G$, the algorithm tries if any vertex in $N_{\prec_\eta}(v_i)\cup N^2_{\prec_\eta}(v_i)$ can form a $k$-plex with $P$. So, this will at most add a factor of $|N_{\prec_\eta}(v_i)|+ |N^2_{\prec_\eta}(v_i)|\leq D+D\Delta$. In total, the running time is in $O((D+D\Delta)\sum_i |N^2_{\succ_\eta}(v_i)|^{k}\gamma_k^{|N(v_i)|})=O(n(D\Delta)^{k+1}\gamma_k^D)$.

\end{proof}

\paragraph{Proof of Prune Rule \ref{prune_size}}
\begin{proof}
Fix $v$ with the leading vertex $v_i$ in Lemma \ref{lemma_size}.
\begin{itemize}
  \item If $u\in N_{\succ_\eta}(v_i)$, then $(u,v_i) \in E$. Thus for any vertex $u \in P$, $|N(u)\cap N(v_i)\cap V_i| \ge |N(u)\cap N(v_i) \cap P| \ge l-2k$,
  \item If $u\in N_{\succ_\eta}^2(v_i)$, then $(u,v_i) \notin E$. Thus for any vertex $u \in P$, $|N(u)\cap N(v_i)\cap V_i| \ge |N(u)\cap N(v_i) \cap P| \ge l-2k+2$.
\end{itemize}
\end{proof}

\paragraph{Proof of Prune Rule \ref{prune_common}}
\begin{proof}
\label{proof_common}
% \color{red}
It is clear $|N_{G_i}(u) \cap N_{G_i}(v) \cap \{v_i\}| + |N_{G_i}(u) \cap N_{G_i}(v) \cap S| + |N_{G_i}(u) \cap N_{G_i}(v) \cap C_s| \ge | N(u) \cap N(v) \cap P|$.
Because $u,v \in S$, then $|N_{G_i}(u) \cap N_{G_i}(v) \cap\{v_i\}|=0$ and $|N_{G_i}(u) \cap N_{G_i}(v) \cap S|\le k-1-2=k-3$. Thus, $|N_{G_i}(u) \cap N_{G_i}(v) \cap C_s| \ge |N(u) \cap N(v) \cap P|-max(k-3,0)$.
According to Lemma \ref{lemma_size}, $|N(u) \cap N(v) \cap P|$ has a lower bound depending on whether $(u,v) \in E$ or not. Combining that, we present Prune Rule \ref{prune_common} as above.
\end{proof}

\end{document}